\documentclass[manuscript,screen,natbib]{acmart}
\usepackage{minted}
\usepackage{subcaption}
\usepackage{xcolor}
\usepackage{circledsteps}
\usepackage{tikz}
\usepackage{multirow}

\usepackage{booktabs}
\usepackage{longtable}
\usepackage{tabularx}
\usepackage{colortbl}
\usepackage{array}
\usepackage{ragged2e}

\newcolumntype{P}[1]{>{\RaggedRight\arraybackslash}p{#1}}
\usepackage[most]{tcolorbox}
\tcbuselibrary{minted,skins,breakable}
\usepackage{listings}

\lstdefinestyle{promptstyle}{
    basicstyle=\ttfamily\scriptsize,
    breaklines=true,
    breakatwhitespace=true,
    columns=fullflexible,
    keepspaces=true,
    showstringspaces=false,
    tabsize=2,
    aboveskip=0pt,
    belowskip=0pt,
    xleftmargin=0pt,
    xrightmargin=0pt,
    frame=none
}

\newtcblisting{promptbox}{
    enhanced,
    breakable,
    listing only,
    colback=white,
    frame hidden,
    boxrule=0pt,
    arc=0pt,
    outer arc=0pt,
    left=0pt,
    right=0pt,
    top=1.5mm,
    bottom=1.5mm,
    before skip=0pt,
    after skip=0pt,
    borderline north={0.4pt}{0pt}{black},
    borderline south={0.4pt}{0pt}{black},
    listing options={
        style=promptstyle
    }
}

\definecolor{rqbg}{HTML}{F3F3F3}      
\definecolor{rqtitle}{HTML}{444444}  
\definecolor{evintro}{HTML}{1B5E20}  
\definecolor{evchange}{HTML}{E65100} 
\definecolor{evremove}{HTML}{B71C1C} 
\definecolor{algcmt}{HTML}{757575}   
\definecolor{gitcmd}{HTML}{4527A0}   
\newtcolorbox{rqsummary}[2][]{%
  enhanced,
  breakable,
  colback=rqbg,
  colframe=white,
  boxrule=0pt,
  arc=0pt,
  left=6mm,
  right=2mm,
  top=2mm,
  bottom=2mm,
  before skip=10pt,
  after skip=10pt,
  overlay={
    \fill[rqtitle] (frame.south west) rectangle ([xshift=5mm]frame.north west);
    \node[rotate=90, anchor=center, text=white, font=\sffamily\small]
      at ([xshift=2.5mm]frame.west) {#2};
  },
  #1
}

\AtBeginDocument{%
  }

\newcommand{\gwt}{\textsc{GWT}}

\newcommand{\bcircle}[1]{\Circled[fill color=black, inner color=white]{#1}}

\begin{document}

\title{An Agent-Based Approach for Automating Behavior-Driven Development}

\author{Luciano Belo de Alcântara Júnior}
\email{luciano.alcantara@dcc.ufmg.br}
\orcid{0009-0003-2390-159X}
\affiliation{%
  \institution{Universidade Federal de Minas Gerais}
  \city{Belo Horizonte}
  \state{Minas Gerais}
  \country{Brasil}
}

\author{Diego Elias Costa}
\email{diego.costa@concordia.ca}
\orcid{0000-0001-7084-2594}
\affiliation{%
  \institution{Concordia University}
  \city{Montréal}
  \state{Québec}
  \country{Canada}
}

\author{João Eduardo Montandon}
\email{joao@dcc.ufmg.br}
\orcid{0000-0002-3371-7353}
\affiliation{%
  \institution{Universidade Federal de Minas Gerais}
  \city{Belo Horizonte}
  \state{Minas Gerais}
  \country{Brasil}
}

\begin{abstract}
  Behavior-Driven Development (BDD) enables stakeholders to express the expected system behavior as a collection of test scenarios in a natural language fashion.
  These scenarios can be automatically executed just like unit tests, ensuring system correctness at the acceptance level.
  However, making these specifications executable requires developers to implement additional code to connect each scenario to the system code they should execute.
  In this paper, we present \textbf{BrinAI}, a multi-agent approach that automates the execution of BDD scenarios by generating the step definitions that links both scenarios and the code of the system under test.
  We leveraged two new datasets of fully executable BDD scenarios mined from open-source projects---\textbf{GWT-Verified} and \textbf{GWT-Light}---to evaluate BrinAI against Large Language Models (LLMs) baselines.
  BrinAI executed 77 out of 171 BDD scenarios (45.0\%), while the strongest baseline did it for 22 (12.9\%);
  the difference remained above 27.3 p.p. regardless the programming language used.
  By contrast, BrinAI presented a significantly higher cost; US\$0.0973 against US\$0.0053 for the baseline.
  Our work discusses the implications of agent-based approaches outperforming single-shot LLM generation, the need for metrics that assess both functional and behavioral success in BDD scenario generation, and the importance of considering execution costs when adopting agent-based solutions.
\end{abstract}




\maketitle

\section{Introduction}

In modern software development, automated testing is crucial for ensuring software quality and supporting continuous delivery~\cite{Wang2022, Shahin2017, Humble2010}.
The benefits of automated testing are well established, including early detection of defects, regression prevention, and documentation of expected behavior~\cite{tdd_fowler, Winters2020, Harman2024, HoraFraser2026}.
Automated tests can verify system behavior at different levels:
unit tests check components in isolation, integration tests verify interactions between these components, and end-to-end tests measure the system behavior at the acceptance level.

Traditionally, most efforts in the literature focus on automating test generation at function level, i.e.,~unit testing~\cite{Godefroid2005,Pacheco2007,Fraser2011}. These efforts have explored different techniques, including symbolic execution, feedback-directed random testing, and search-based test generation. Such approaches typically rely on program structure, execution feedback, and code coverage criteria to automatically generate test cases.
While unit tests have been extensively studied, the generation of system-level testing remains underexplored~\cite{Meglio2025}.

In this context, Behavior-Driven Development (BDD) emerged as a promising approach to bridge the gap between high-level requirements and executable tests~\cite{Wynne2017}.
In BDD, people can express expected system behavior as a set of scenarios, all written in a semi-structured natural language format, called Gherkin.
Each scenario is written as a sequence of steps---typically in \texttt{Given--When--Then} format---that describe its initial context, the actions needed to fulfill it, and the expected outcome.
Developers can then parse these semi-structured scenarios and implement the corresponding code to be executed for each step, called \textit{step definitions}.
This implementation will make the scenarios executable, thus enabling the system to be automatically verified at the acceptance level.

Ultimately, \textbf{BDD allows stakeholders to express system behavior in a human-readable format, while enabling the execution of these tests automatically and ensure correctness at system level}.
However, the adoption of BDD largely depends on developers ability to implement the step definitions that connect the scenarios to the source code of the system under test~\cite{Binamungu2018}.
Any automation of this process implies several challenges, including the broad scope of system-level behaviors spanning multiple interacting components~\cite{Wiecher2021}, the difficulty of translating high-level behavioral specifications into concrete test operations, and the need to identify appropriate execution conditions and test data to reliably verify the expected behavior~\cite{Wang2022}.

Large Language Models (LLMs) have been used in several software engineering tasks, including code generation~\cite{du2024classeval,liu2023evalplus,wang2023codet5plus}, software documentation~\cite{guelman2026quality,geng2024summarizers,sun2025summarization}, software migration~\cite{trabalho_altino_ast,trabalho_aylton,trabalho_nadi_islam,Islam_2024,bairi2024codeplan}, and unit test generation~\cite{lemieux2023codamosa,yuan2024chattester,Schafer2024}.
Despite the success reported in many of these tasks, repository-level software engineering problems remain challenging for direct LLM-based generation. Relevant information is often scattered across multiple files, making it difficult to identify and incorporate the context required for a correct implementation~\cite{jimenez2024swebench,Liu2024RepoBench,Zhang2023RepoCoder,Wang2025CodeRAGBench}.
Furthermore, generated code may fail to compile or execute successfully, and its functional correctness cannot be established from syntactic plausibility alone~\cite{Schafer2024,Harman2024}.
Recent approaches have therefore explored structured workflows that combine repository exploration, code generation, and execution-based validation and refinement~\cite{xia2024agentlessdemystifyingllmbasedsoftware,yang2024sweagentagentcomputerinterfacesenable,CoverUp2025}.

In this work, we propose \textsc{BrinAI}, a multi-agent approach to generate step definitions from BDD scenarios and the system code of the project under test.
BrinAI decomposes the step generation task into six parts:
(a) \textit{BDD snapshot} provides the target scenario, project source code, and execution configuration;
(b) \textit{Plan \& contextualize} analyzes the scenario and project context to establish an implementation plan aligned with the available dependencies and testing framework;
(c) \textit{Generate} produces step definitions guided by the plan and relevant project code;
(d) \textit{Evaluate} performs local validation and, when these checks pass, executes the tests in an isolated environment to collect execution feedback;
(e) \textit{Correct \& rank} uses validation and execution feedback to propose repairs, ranks candidate implementations using local validation signals, and selects a candidate
for re-evaluation; and
(f) \textit{Terminal outcome} reports success or the reason for termination, together with the best available implementation and its evaluation results.
Essentially, BrinAI combines contextual information of the system under testing and execution feedback from running the test under automation.

We also introduce two new BDD datasets to support the evaluation of BrinAI and other approaches for generating step definitions: \textsc{GWT-Verified} and \textsc{GWT-Light}.
The former was built upon the GivenWhenThen dataset~\cite{GivenWhenThenDataset}, and gathers fully executable BDD scenarios from open-source projects, comprising of 713 executable BDD scenarios, containerized so that they can be executed in a controlled environment.
The latter is a subset of 171 snapshots from \textsc{GWT-Verified}, selected using stratified sampling to ensure a representative distribution of programming languages, project domains, and scenario complexity~\cite{Nagappan2013Diversity}.

We evaluate BrinAI on the GWT-Light dataset, comparing its effectiveness in terms of successfully executed scenarios, its efficiency in terms of number of tokens processed, and its superiority over baselines based on direct generation with LLMs.
Overall, BrinAI was able to successfully execute 77 out of 171 scenarios (45.0\%).
The agent performed best on Java and JavaScript ecosystems, with 52.3\% and 55.3\% successful executions, respectively;
Ruby was the most challenging one, with 37.1\% of scenarios executed successfully.
The average cost of executing BrinAI was US\$ 0.0973 per scenario, with input and output tokens remaining approximately at 426K and 33K, respectively; an input-to-output ratio of 12.7.
BrinAI consistently outperforms LLM baselines across all three ecosystems evaluated: 52.3\% vs 25.0\% for Java, 55.3\% vs 23.7\% for Node.Js, and 37.1\% vs 3.4\% for Ruby.

\noindentparagraph{Contributions.}
We highlight the following contributions of this work:

\begin{enumerate}
  \item \textbf{BrinAI}, the first agent approach for generating BDD step definitions for multiple languages. BrinAI combines repository exploration, selective context retrieval, code generation, and iterative repair guided by test execution feedback.

  \item \textbf{GWT-Verified} and \textbf{GWT-Light}, two datasets containing independent, containerized, and executable BDD snapshots extracted from real-world open source projects.
        In their current version, these datasets contain 713, and 171 snapshots, respectively.
        To the best of our knowledge, these are the first datasets to provide a reproducible container-based environments for executing and evaluating Cucumber BDD scenarios, with ready-to-use execution commands.

  \item \textbf{A comprehensive evaluation benchmark}, where we compared the effectiveness and cost of BrinAI against LLM baselines over the GWT-Light dataset.
\end{enumerate}

\noindentparagraph{Paper organization.}
The remainder of this paper is organized as follows.
Section \ref{sec:background} introduces what BDD is, its needed artifacts, and discusses the challenges of employing it in software development.
Section \ref{sec:gwt} describes the whole process to build both GWT-Verified and GWT-Light datasets.
The architecture of BrinAI is provided next in Section \ref{sec:brinai}.
Sections \ref{sec:evaluation} and \ref{sec:results} present the research questions we leveraged to evaluate BrinAI and the results we obtained, respectively;
some implications of our findings are discussed in Section \ref{sec:discussion}.
Sections \ref{sec:threats} and \ref{sec:related-work} enumerate threats to validity and related work, respectively.
Finally, Section \ref{sec:conclusion} concludes the paper and pinpoints some future directions.

\section{Background}
\label{sec:background}
\subsection{Behavior-Driven Development}

Behavior-Driven Development (BDD) is a software development approach that promotes collaboration between technical and non-technical stakeholders by describing the expected behavior of a system through scenarios expressed in a language shared by the development team and domain experts~\cite{Wynne2017}. These scenarios capture business rules and acceptance criteria in a form that can be understood by different stakeholders and connected to automated tests.

While Test-Driven Development (TDD) primarily guides software development through tests written at the implementation level, \textbf{BDD emphasizes specifying the expected behavior of the system from the perspective of its stakeholders}. In BDD frameworks such as Cucumber, these specifications are commonly expressed using Gherkin, a semi-structured language designed to be readable by both humans and automated tools.
In this language, each scenario consists of a sequence of steps describing an initial context, an action or event, and an expected outcome, typically represented by the \texttt{Given}, \texttt{When}, and \texttt{Then} keywords, respectively.

By adopting this approach, BDD addresses two main challenges in software development. First, it bridges the communication gap between technical and non-technical stakeholders by establishing a shared language based on concepts from the application domain~\cite{Evans2003}. Second, its semi-structured specifications can be connected to executable code, enabling automated tools to verify whether the implemented system exhibits the described behavior~\cite{Wynne2017}.
In practice, when these scenarios are maintained and continuously executed against the system, they can serve as living documentation that evolves together with the software and reflects its current expected behavior.

BDD's relevance in industry is well-established.
Cucumber, its most widely used implementation, has been ported to over a dozen languages and ecosystems (e.g., SpecFlow for .NET, Behave for Python, and Cucumber-JVM for Java), making it one of the major standards for acceptance testing in industry.
This adoption is not restricted to proprietary software.
\citet{Zampetti2020} mined a large sample of open-source repositories from GitHub and found that Cucumber-based BDD scenarios are present in a wide range of project domains, with teams reporting that keeping feature files and their automated tests synchronized was a primary motivation for adoption.

\subsection{\textbf{The BDD Testing Stack}}
\label{sec:bdd-test-stack}

To establish an automated BDD testing stack, developers have to bridge the test scenario written in Gherkin with the code that executes the test.
To achieve this, developers implement a set of \textit{step definitions}: code that maps each step in the scenario to its corresponding implementation.
Step definitions are implemented in the same programming language as the system under test, and they call the API provided by the \textit{system code} to execute the actions expected for the step.
In other words, creating a comprehensive set of BDD test scenarios requires at least three main components: (a) the \textit{feature files} containing the test scenarios, (b) the \textit{steps definition files} implementing the behavior of the steps described in the scenarios, and (c) the \textit{system code files} containing the implementation of the system under test.

\begin{figure*}[t!]
  \centering
  \includegraphics[width=0.9\textwidth]{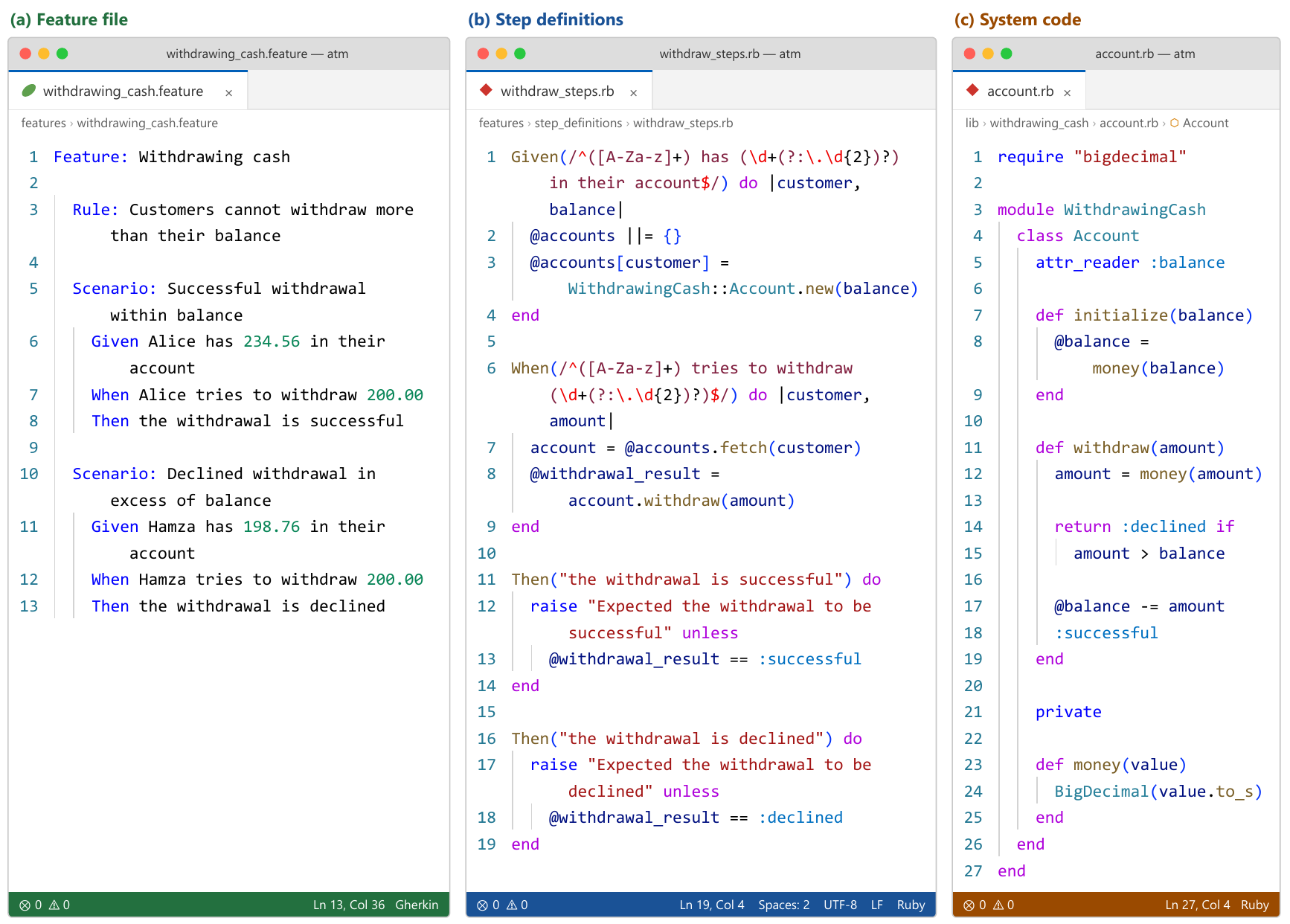}
  \caption{A complete BDD example: the feature file (a) is parsed by the step
    definitions (b), which executes the Ruby \texttt{Account} implementation (c).}
  \Description{GWT-Verified pipeline.}
  \Description{Three side-by-side code panels illustrating a cash withdrawal
    BDD example: a Gherkin feature file, its Ruby step definitions, and the
    Ruby Account implementation, connected by directional arrows.}
  \label{fig:withdraw-example}
\end{figure*}

\noindentparagraph{A Complete BDD Example.}
Figure \ref{fig:withdraw-example} illustrates these three components together with a complete example.
The feature file was extracted from the official Cucumber documentation,\footnote{\url{https://cucumber.io}} while the step definitions and the system code are one possible implementation of this scenario in Ruby.

Figure~\ref{fig:withdraw-example}(a) declares the ``Withdrawing cash'' feature and the rule that customers cannot withdraw more money than their account balance.
The feature file contains two scenarios that exercise this rule.
In the first one, it gives Alice a balance of 234.56 and expects a withdrawal of 200.00 to succeed;
the second gives Hamza a balance of 198.76 and expects a withdrawal of 200.00 to be declined.
In both scenarios, the steps describe the account state, the attempted action, and the expected result with the Given--When--Then structure.

Figure~\ref{fig:withdraw-example}(b) implements four step definitions connecting the feature file to the system implementation.
The first definition extracts the customer's name and balance from the Given step, initializes an account, and stores it in the shared \texttt{@accounts} list.
The next one extracts the customer and withdrawal amount, retrieves the corresponding account, and calls its \texttt{withdraw()} method, storing the returned result in \texttt{@withdrawal\_result}.
The last two definitions validate the result by raising an error when it does not match the expected \texttt{:successful} or \texttt{:declined} outcome. Thus, the step definitions translate the natural-language steps into software updates, system calls, and assertions.

Finally, Figure~\ref{fig:withdraw-example}(c) presents the \texttt{WithdrawingCash::Account} class with the behavior verified by the steps.
The class stores the account balance and provides a \texttt{withdraw()} method to perform the withdrawal operation.
The method subtracts the requested amount from the balance and returns \texttt{:successful} if there are sufficient funds; otherwise, it returns \texttt{:declined}.

When combined, these files provide a comprehensive view of the presented BDD test scenario.
The \textit{feature file} specifies, in human-readable Gherkin, the expected outcomes for withdrawals within and beyond a customer's balance.
The \textit{step definitions file} bridges the natural-language steps and the executable code by creating accounts, invoking the withdrawal operation, and checking its result.
Finally, the \textit{system code file} implements the rules that determines whether each withdrawal is successful or declined for a given account.
This complete structure ensures that the BDD test scenario can be fully automated and executed as intended.

\subsection{The Step Definition Bottleneck}
\label{sec:step-def-bottleneck}
Although the BDD testing stack cleanly separates feature files, step definitions, and system code, in practice step definitions are costly to produce and maintain.
Every new Gherkin step requires a corresponding hand-written method connecting the scenario to the system's API.
In a survey of 75 BDD practitioners from 26 countries, \citet{Binamungu2018} report that developers frequently face a steep learning curve and struggle to apply BDD correctly.
\textbf{The growth of feature files and step definitions has been highlighted as a main obstacle to maintaining BDD test suites}, increasing development time and costs.

This suggests that automating, even partially, the generation of step definitions from feature files and system code is a promising direction to reduce the manual effort required to adopt and sustain BDD in practice, an opportunity we investigate in this paper.


\section{The GivenWhenThen Dataset (GWT)}
\label{sec:gwt}

To evaluate our agentic approach, which supports partial or complete automation of BDD step definitions, we need a comprehensive dataset of BDD artifacts.
In a previous work, we have introduced the \gwt\ dataset, which contains 2,289 complete BDD test scenarios collected from 756 open-source projects~\cite{GivenWhenThenDataset}.
In short, the dataset captures the three main components of the BDD testing stack described in Section~\ref{sec:bdd-test-stack}---feature files, step definitions, and system code---and compiles them into a structured format preserving the content and relationship between these artifacts.

Figure \ref{lst:json-example} shows an example of a BDD test scenario stored in \gwt.
In this record, one can find the name of the repository from which the scenario was extracted, the programming language used to implement the step definitions and system code files, and the paths to the feature file, step definitions file, and system code files.

\begin{figure}[ht]
  \begin{minted}[
    frame=lines,
    framesep=2mm,
    breaklines,
    autogobble,
    linenos,
    fontsize=\small,
  ]{json}
{
  "repository": "Test-Architect/playwright-java-bdd-sample",
  "language": "java",
  "normalized_repo_path": "<path>/test-architect_playwright-java-bdd-sample",
  "feature_file": "src/test/resources/features/google_search.feature",
  "feature_content": "Scenario: Search for Playwright on Google Given I am...",
  "step_definitions_file": "<path>/steps/GoogleSearchSteps.java",
  "step_definitions_content": "... public class GoogleSearchSteps { private ...",
  "system_code_files": [
    {
      "name": "GoogleHomePage.java",
      "path": "<path>/framework/pages/GoogleHomePage.java",
      "content": "... public class GoogleHomePage { private final Page page ..."
    }
  ]
}
  \end{minted}
  \caption{A JSON example containing a BDD test scenario stored in GWT.}
  \Description{Example JSON object representing a BDD test scenario, including
    the repository name, programming language, feature file and its content,
    step definitions, and associated system code files.}
  \label{lst:json-example}
\end{figure}

Despite the comprehensive nature of the GWT dataset, its construction process did not consider the executability of the test scenarios.
Specifically, the presence of BDD artifacts do not guarantee that the scenarios can be executed since these scenarios actually depend on the whole project properly configured, set up, and running in a specific environment.

\subsection{The GWT-Verified Dataset}
\label{sec:gwt-verified}

To address this gap, we introduce \textbf{GWT-Verified}, a new dataset derived from GWT that ensures each test scenario can be executed in an isolated and preconfigured environment.
In essence, GWT-Verified provides a complete snapshot of the project for each scenario, enabling its execution on any machine without additional setup.

\begin{figure*}[t!]
  \centering
  \includegraphics[width=1\textwidth]{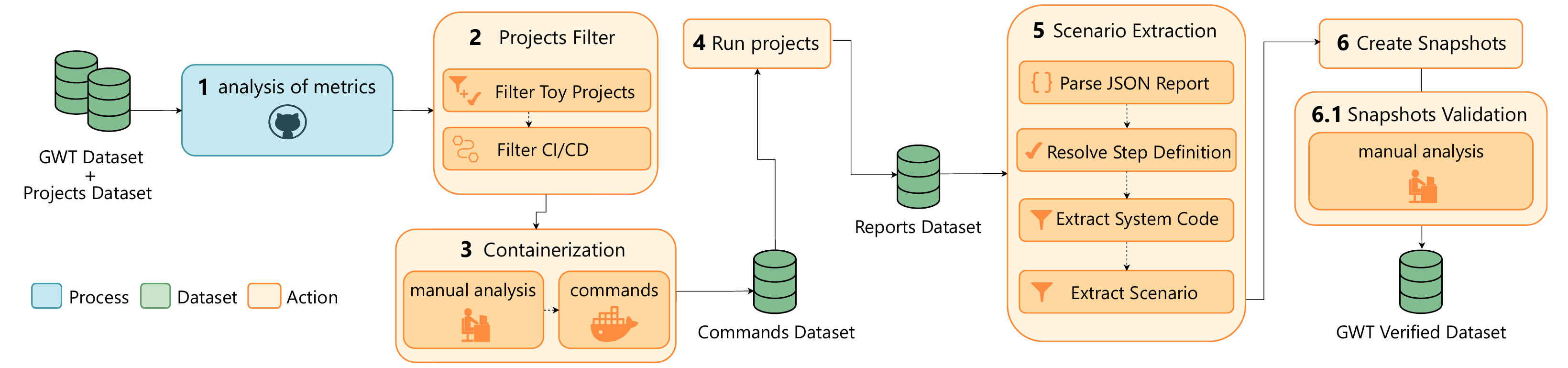}
  \caption{GWT-Verified pipeline.}
  \Description{GWT-Verified pipeline.}
  \label{fig:pipeline-gwt-verified}
\end{figure*}

Figure \ref{fig:pipeline-gwt-verified} depicts the steps taken to create these reproducible scenarios.
We start by analyzing a set of metrics from the initial corpus to filter out projects that are less likely to be executable.
Then, we manually containerize the remaining projects and standardize their execution commands.
Next, we execute the projects and consolidate the Cucumber reports to identify which scenarios were successfully executed.
Finally, we extract the structured information for each scenario and create a verified snapshot of the project, ensuring that it can be executed in isolation.
These steps are described in detail next.

\noindentparagraph{\bcircle{1} Metrics Analysis.}

As the first stage, we identified characteristics and attributes that could help us detect the projects with higher chances to be executed.
Based on prior work~\cite{munaiah2017curating, kalliamvakou2014promises}, we analyzed them according to four perspectives, as described next.

\begin{itemize}
  \item \textit{Repository Activity.}
        We collected projects meta-data to determine their maturity level.
        We collected the number of stars, forks, watchers, open issues, creation and recent activity dates, commits, contributors, branches, releases, default branch, and archive status.
  \item \textit{Automation.}
        We verified the presence of CI/CD configuration files, e.g.,~\texttt{config.yml}, \texttt{action.yml}, \texttt{.travis.yml}.
        Supposedly, the presence of these files indicate their projects are configured to be executed in different environments.
  \item \textit{BDD specification.}
        We counted the number feature files present in each repository as an indicator of BDD adoption in each one.
  \item \textit{Structure.}
        We calculated the total number of repository files, total number in the inferred primary language, and the size in bytes.
        The file count excluded dependency, build, and cache directories, making the metric closer to the content maintained by the project.
\end{itemize}

\noindentparagraph{\bcircle{2} Projects Selection.}

At this stage, we filtered projects with a higher likelihood of successful execution.
From the 756 projects initially selected from GWT, we first excluded 656 classified as \textit{toy projects}, i.e.,~those whose name contained the terms \textit{practice}, \textit{demo}, \textit{example}, \textit{sample}, \textit{tutorial}, and \textit{bdd}, etc~\cite{munaiah2017curating, kalliamvakou2014promises}.

Another 60 projects did not contain CI/CD artifacts and were discarded.
We relied on these files as a \textit{proxy} for project executability and maturity, since continuous integration or delivery pipelines normally depend on automated procedures for dependency installation, system configuration, and test execution.
After applying this filtering procedure, we end up with 40 projects that are potentially executable.

\noindentparagraph{\bcircle{3} Projects Containerization.}

Having a CI/CD pipeline is not sufficient for reproducing a given BDD test scenario as it can depends on others external configurations, such as dependencies, databases, browsers and third-party APIs.
Therefore, we also manually analyzed each selected repository in order to create virtual environments to execute the test scenarios available in each one~\cite{boettiger2015docker}.

To this end, we examined the directory structure, build configuration, dependency manifests, and container orchestration files.
For each repository, we manually implemented docker artifacts (both Dockerfile and docker-compose.yml) that reproduce the sequence of commands required to execute each one;
we made adjustments---e.g.,~resolving dependency issues, configuring browsers for headless execution, adapting database connection settings, and configuring test runners to generate execution summaries---when needed.
We were able to successfully containerize 27 of the 40 selected repositories.
Most invalid cases failed due to incomplete commands or commands dependent on proprietary credentials/infrastructure and nondeterministic environments.

\noindentparagraph{\bcircle{4} Projects Execution.}

In the fourth stage, we executed each project and analyzed the test results to verify whether the BDD scenarios could be correctly executed in the isolated environment.
We executed Cucumber in each project and collected its generated reports, a JSON file that records the execution of each scenario, including the steps executed, and the test result (e.g.,~\texttt{passed}, \texttt{failed}, or \texttt{skipped}).
Then, for each project, we created a reference file containing basic execution information, such as the programming language, the local repository path, and the path of the Cucumber report generated during test execution.
All this information is used in the next step to precisely extract the BDD scenarios and their associated code.

\noindentparagraph{\bcircle{5} Scenario Extraction.}

The fifth stage unwraps each Cucumber report into a set of scenario instances, i.e.,~the candidates for our dataset. Each instance contains an executed scenario, its step definitions, and the source code associated with those definitions.
This procedure follows four steps, described below.

\noindentparagraph{\textbf{5.1.} Parse Cucumber Report.}
We locate each \textit{feature file} based on the reference provided in the Cucumber report.
Some scenarios use \texttt{Background} blocks to share initial context for test execution;
we associated these blocks to the scenarios that depend on them.
Finally, scenarios without any executed step were discarded.

\noindentparagraph{\textbf{5.2.} Resolve Step Definition.}

In this step, we located the corresponding step definition files used by Cucumber during the execution of each scenario.
This is done by analyzing the \texttt{match.location} property provided by the report for each recorded step.
We employed different resolution procedures depending on the language and the reference format.
For example, in Ruby and Node.js/TypeScript, the reference normally contains the file path and line number, while in Java it may also contain a fully qualified method name, such as \texttt{com.example.steps.MySteps.method(String)}.

\noindentparagraph{\textbf{5.3.} Extract System Code.}
After locating the step definition files, we statically analyze their references to collect the associated system code.
In Node.js, we considered \texttt{require} calls and \texttt{import} statements with relative paths.
In Ruby, we considered \texttt{require}, \texttt{require\_relative}, and CamelCase constants compatible with autoload mechanisms.
In Java, we consider imports from local packages and discards references to known external libraries.

\noindentparagraph{\textbf{5.4.} Extract Scenario.}

In this step we split each feature file into multiple single and independent scenarios, so they can be executed in isolation.
Background blocks are copied to each scenario that depends on them, and scenarios without at least one \texttt{Given} or \texttt{When} are discarded.
In case the feature file contains \texttt{Scenario Outlines}, we generate a distinct scenario for each row in the \texttt{Examples} table, interpolating the values into the scenario name and step texts.
Finally, we excluded duplicates based on the repository, feature file path, and the content of the isolated scenario, keeping only those cases marked as \texttt{passed} according the Cucumber JSON report.

After removing 3 duplicates, 9 scenarios without any \texttt{Given} or \texttt{When} step, and 10 scenarios from automatically generated feature files, 818 scenario instances remained.
From these, we discarded another 82 because their status in the Cucumber JSON report was \texttt{not passed}.
Each of the \textbf{736 scenario instances} contains the scenario, its associated step definitions, and the system code referenced by them.
Table \ref{tab:dataset-curado} shows the number of the extracted scenarios by programming language.
Ruby-based ones had the highest pass rate (98.2\%), followed by Java (88.8\%) and Node.js (78.1\%).

\begin{table}[htbp]
  \centering
  \caption{Instances extracted by language after curation based on execution outcome.}
  \label{tab:dataset-curado}
  \begin{tabular}{lccc}
    \toprule
    \textbf{Language} & \textbf{Extracted scenarios} & \textbf{Passed instances} & \textbf{Pass rate} \\
    \midrule
    Ruby              & 379                          & 372                       & 98.2\%             \\
    Java              & 197                          & 175                       & 88.8\%             \\
    Node.js           & 242                          & 189                       & 78.1\%             \\
    \midrule
    \textbf{Total}    & \textbf{818}                 & \textbf{736}              & \textbf{90.0\%}    \\
    \bottomrule
  \end{tabular}
\end{table}

\noindentparagraph{\bcircle{6} Snapshot Construction and Validation.}

Here we created a snapshot for each scenario extracted previously, containing the isolated scenario, the associated step definitions, and the system code referenced by them.
We start by creating a new directory representing the snapshot, named according to the scenario name, the feature file path, and the SHA-1 hash of the isolated scenario.
We then copy the artifacts needed to execute the scenario: a feature file with the isolated scenario, the associated step definitions, and the system code.
When step definitions are distributed across multiple files, we merge them into a one to reduce interference from unused steps and avoid name conflicts.
Finally, we revisited each scenario and executed them manually to identify possible failures, e.g., erratic, invalid, or semantically inappropriate scenarios.
From this analysis, we removed 23 scenarios (3.1\%).

\noindentparagraph{The GWT-Verified Dataset.}
Table \ref{tab:gwt-verified} presents the distribution of the 713 scenarios that passed the validation process across the three programming language ecosystems.
Ruby-based snapshots account for 51.6\% (368) of the dataset, while Node.js and Java account for 24.8\% (177) and 23.6\% (168), respectively.
Unlike a catalog of plain text feature files mined from public repositories, \textbf{all scenarios in GWT-Verified dataset are fully executable and reproducible in any environment}.
We hope this dataset be useful for researchers and practitioners interested in evaluating BDD test generation and maintenance techniques, as it provides a reliable and diverse set of scenarios that can be executed in a controlled environment.

\begin{table}[htbp]
  \centering
  \caption{Composition of the GWT-Verified dataset by language.}
  \label{tab:gwt-verified}
  \begin{tabular}{lcc}
    \toprule
    \textbf{Language} & \textbf{Snapshots} & \textbf{Share} \\
    \midrule
    Ruby              & 368                & 51.6\%         \\
    Node.js           & 177                & 24.8\%         \\
    Java              & 168                & 23.6\%         \\
    \midrule
    \textbf{Total}    & \textbf{713}       & \textbf{100\%} \\
    \bottomrule
  \end{tabular}
\end{table}

\subsection{Creating a Lightweight Subset of GWT-Verified}
\label{sec:gwt-light}

Despite the comprehensive nature of GWT-Verified, its number of scenarios can impose a significant computational cost for LLM and agent-based evaluations.
Other studies circumvent this problem by randomly selecting a sample from the original dataset ~\cite{bouzenia2025repairagent,gou2024critic,ma2025thinking}, but this approach can introduce a selection bias, thus compromising the representativeness of the evaluation~\cite{Nagappan2013Diversity}.
We tackle this by generating a subset that preserves the diversity of the original dataset while reducing its size, which we call \textbf{GWT-Light}.

Specifically, we applied the coverage-based sample selection strategy proposed by \citet{Nagappan2013Diversity}.
This approach performs a systematic stratified sampling of the original dataset based on a set of dimensions that represent the structural diversity of the data to be sampled.
In our case, we defined a set of BDD-specific dimensions that capture the structural characteristics of the scenarios and their associated code.
For example, we leveraged the number of steps, the presence of \texttt{Background} and \texttt{Scenario Outline} blocks, the number of associated step definitions (e.g.,~\texttt{Given}, \texttt{When}, \texttt{Then}), and the number of system code files referenced by the step definitions.
We also considered general repository metadata, such as programming language and project size.
The selection procedure considered 25 dimensions: 13 at the repository level and 12 at the scenario level, to ensure that the selected subset preserves the diversity of the original dataset.

\noindentparagraph{\textbf{The GWT-Light Dataset.}}

The final GWT-Light dataset contains \textbf{171 snapshots}---a reduction of \textbf{76.02\%} compared to the complete one---while preserving \textbf{96.30\%} of the diversity of the original dataset.
Table \ref{tab:gwt-light-composition} summarizes the characteristics of the GWT-Light dataset.
As expected, the distribution of snapshots by programming language remained similar, with Ruby accounting for 52.1\% (89), Node.js for 22.2\% (38), and Java for 25.7\% (44).

\begin{table}[htbp]
  \centering
  \caption{Composition of the GWT-Light dataset by language.}
  \label{tab:gwt-light-composition}
  \begin{tabular}{lrrrr}
    \toprule
    \textbf{Language} & \textbf{Snapshots} & \textbf{Share}    \\
    \midrule
    Java              & 44                 & 25,73\%           \\
    Node.js           & 38                 & 22,22\%           \\
    Ruby              & 89                 & 52,05\%           \\
    \midrule
    \textbf{Total}    & \textbf{171}       & \textbf{100,00\%} \\
    \bottomrule
  \end{tabular}
\end{table}


\section{BrinAI}
\label{sec:brinai}

\subsection{System Overview}
\label{sec:visao-geral}

\textbf{BrinAI is a multi-agent system specialized in generating BDD step definitions}. The initial motivation is the bottleneck of frameworks such as Cucumber~\cite{10.1007/978-3-642-30561-0_19}.
As each BDD scenario requires a precise and minimally correct manual implementation before being executed, this task is generally delegated to developers who know both the requirements and system code needed to implement the steps~\cite{yang2025requirementsbasedtestgenerationcomprehensive,381473.381476}.

The agent input is a \textbf{snapshot} of the repository with (a) its target language, (b) its \textbf{feature} file, (c) the target scenario, and (d) its source code.
The output is a complete and executable \textbf{step definitions file} or, in case of failure, a structured report of the error and the trajectory.

\begin{figure}[htbp]
  \centering
  \includegraphics[width=\textwidth]{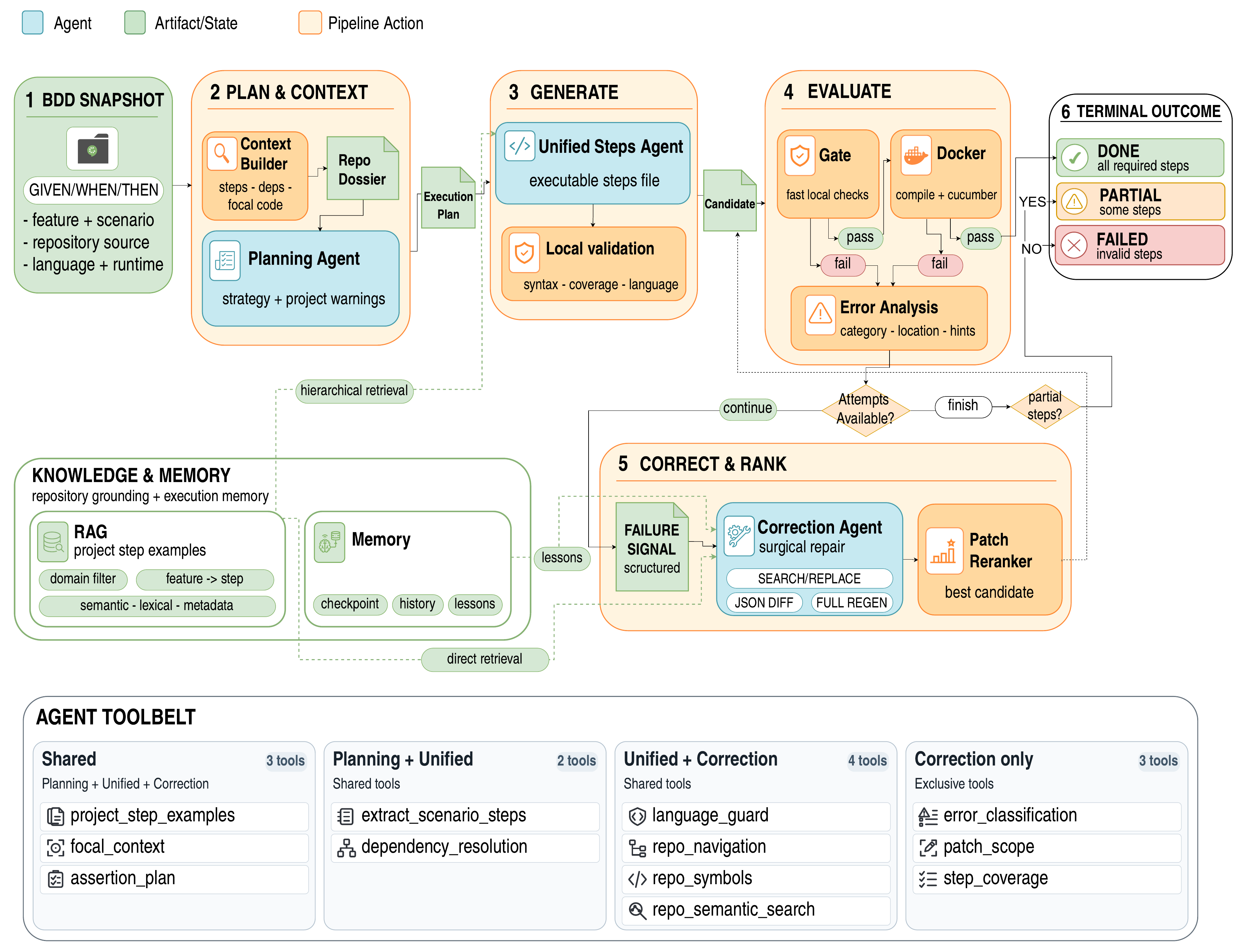}
  \caption{BrinAI agentic loop, inspired by the Plan-Execute-Evaluate-Correct paradigm \Description{Wang_2024,yang2024sweagentagentcomputerinterfacesenable}. The PlanningAgent produces the ExecutionPlan consumed by the UnifiedStepsAgent; the evaluator layer --- gate, Docker, Error Analysis, and reranker --- drives the correction loop until the tests pass or the budget is exhausted.}
  \label{fig:pipeline}
\end{figure}

The agent loop operates with three distinct subagents, inspired by the Plan-Execute-Evaluate-Correct paradigm, a strategy that outperforms single-prompt approaches in software engineering tasks~\cite{yang2024sweagentagentcomputerinterfacesenable, hong2024metagptmetaprogrammingmultiagent}.
Each agent follows a ReAct architecture~\cite{yao2023reactsynergizingreasoningacting}, which autonomously decides when it interacts with its environment.
A state graph orchestrates the roles deterministically, which guarantees predictability and auditability at the macro level, while each agent retains autonomy over its own reasoning process.
The \textit{PlanningAgent} reads the snapshot, constructs the deterministic context, and produces the generation plan.
The \textit{UnifiedStepsAgent} writes the complete step definitions file and submits it for evaluation.
The \textit{CorrectionAgent} repairs failed code through incremental patches or complete regeneration.
An evaluator layer validates each candidate and decides whether the loop continues, escalates, or terminates.
Each agent has well-defined input and output contracts, and agents communicate only through the shared state.
Figure~\ref{fig:pipeline} illustrates the complete pipeline described above. The next sections detail each agent in the pipeline.

\subsection{PlanningAgent}
\label{sec:planningagent}

\subsubsection{Role and design principle.}

The PlanningAgent is the first agent in the loop, and its role is to transform the raw repository snapshot into a rich and structured Execution Plan.
Its central design principle is a clear separation between deterministic and LLM work~\cite{wei2023chainofthoughtpromptingelicitsreasoning, liu2021pretrainpromptpredictsystematic,huyen_ai_2025}.
Everything that can be extracted directly from the repository---e.g.,~language and runtime detection, dependency resolution, extraction of scenario steps, creation of the codebase context, etc---is computed deterministically and assembled into a structured dossier.
The LLM is then invoked for decisions that require reasoning: to formulate the implementation strategy and list project-specific pitfalls.

\subsubsection{Agentic reasoning and tools.}

The model operates as a ReAct~\cite{yao2023reactsynergizingreasoningacting} agent and has access to a set of tools for plan construction, including (a) step extraction, (b) dependency resolution, (c) project step examples, (d) focal context, and (e) an assertion plan.
The focal context tool extracts relevant terms from the scenario name and steps, ranks system code snippets based on their relevance to these terms and keep the highest-ranked ones, following an approach similar to code search~\cite{cambronero2019deeplearningmetcode}.

The step examples scan existing step definitions in the project and summarize local style, common signatures, and hook patterns. We pay special attention to exclude the target step definition file from the scanning process to prevent data contamination.
The agent autonomously decides which tools to invoke according to the available context, and the telemetry from each call is recorded for subsequent analysis.

\subsubsection{Output Format}

The agent receives the required project artifacts and returns a JSON object with two fields: \texttt{strategy}, which briefly describes the proposed implementation approach, and \texttt{warnings}, which identifies potential project-specific issues. When \emph{Background} steps or shared state across scenarios are present, the prompt requires the strategy to address setup and teardown using \texttt{Before}/\texttt{After} hooks and idempotent steps\footnote{A step is idempotent when multiple executions produce the same final state as a single execution. This matters because \emph{Background} steps run before each scenario in a feature; a non-idempotent step (e.g., a direct \texttt{INSERT} instead of an \texttt{upsert}) may fail or leave residual state across scenarios. State dependencies are among the documented causes of \emph{flaky} tests \cite{luo2014empirical}.}. After the agent's response, the pipeline augments the resulting \texttt{ExecutionPlan} with repository memory, including information about approaches that failed in previous snapshots of the same repository during the current execution. This information is provided as additional context for the subsequent generation step.

\subsection{UnifiedStepsAgent}
\label{sec:unifiedstepsagent}

\subsubsection{Roles and design principles}
The UnifiedStepsAgent is responsible for the production of a syntactically valid step definitions file. It currently supports TypeScript, JavaScript, Ruby, and Java through a unified generation pipeline with language-specific adaptations \cite{feng2020codebertpretrainedmodelprogramming, Chen2021EvaluatingLL}.
The pipeline selects the language-specific rule set used to detect the libraries required for test execution (HTTP client, assertion library, etc.).
From this information, the agent constructs the system prompt, including the persona (role) and target framework, the output contract, the language ecosystem rules, and the allowed dependencies.

In turn, the user message\footnote{The user message refers to the task-specific message dynamically constructed by BrinAI and sent to the underlying LLM under the \texttt{user} role, rather than a message directly written by an end user.} organizes the context sections from the strongest constraints (the steps to implement, the libraries, the requirement to use the actual system APIs rather than mocks, and the treatment of DataTable/DocString and Background) to auxiliary sections, which, when RAG is active, comprise local examples, system code, retrieved context, and the complete feature, in accordance with the way language models handle information in long prompts \cite{liu2021pretrainpromptpredictsystematic}.
As a ReAct agent, the generator invokes its tools to confirm dependencies, examples, and assertion plans before it writes the file.

\subsubsection{Local validation}

The generated steps are validated across multiple dimensions. As shown in the figure, local validation immediately follows \textit{Generate}. The \texttt{UnifiedStepsAgent} checks the output for conversational text, remaining code fences or markers, and signs of truncation. It also checks whether each mandatory step matches exactly one definition and whether any step pattern is duplicated. If a check fails, the agent uses the diagnostic to request a revised file.

Once the candidate passes local validation, it is forwarded to the pre-execution gate, which applies deterministic language-specific repairs. For Ruby and JavaScript, the gate also checks syntax using the corresponding parser when available (\texttt{ruby -c} and \texttt{node --check}, respectively).

\subsubsection{Gate, Docker, and Error Analysis.}

The pre-execution gate checks each candidate step-definition file before execution. For Ruby and JavaScript, the gate uses the corresponding language parser, when available locally, to check whether the candidate conforms to the grammar of the target language without executing the code. For TypeScript and Java, it applies language-specific rules, such as checking for unmatched delimiters in TypeScript and verifying that a Java public class name matches the target filename. To check pattern coverage, the gate extracts the patterns declared in the candidate file and verifies that each expected Gherkin step matches exactly one distinct pattern.

If a blocking check fails, the gate returns a structured failure reason to the correction loop. This provides feedback before Docker execution and avoids running candidates that fail these checks.

Approved candidates are forwarded to a Docker container for execution-based evaluation~\cite{Chen2021EvaluatingLL}. The containerized run determines whether the candidate passes. It compiles the project when required and executes Cucumber with the generated step definitions. If the candidate causes an error, the logs are forwarded to Error Analysis, which matches them against known error patterns and sends the error category, diagnostic hints, and recommended fixes to the correction loop. We adopted this error-handling strategy from similar approaches in the literature~\cite{Just2014, 381473.381476}.

\subsection{CorrectionAgent}
\label{sec:correctionagent}

\subsubsection{Role and design principle.}

We designed the \textit{CorrectionAgent} to, given a failure, generate incremental patches to fix it as simple as possible~\cite{zhang2025patchempoweringlargelanguage}.
The CorrectionAgent is invoked when Docker reports an execution failure that requires a repair to the step definitions file.
The prompt instructs it to use the error log to identify the cause of the failure and propose the smallest change that fixes it while preserving working code.
When the pipeline requests a complete file, e.g., because the current file is truncated, the prompt instructs the agent to return the entire corrected file while preserving valid step definitions.

\subsubsection{Correction memory sources}

The CorrectionAgent stores past failed attempts in memory, consolidates the failure information from the Pipeline, and injects it into the prompt to guide correction.
Examples are the current code that failed, the Docker log, the failure location, the deduplicated history of all attempts with their lessons learned, and a summary of the generator's own trajectory (the strategy it followed and how many times it has already failed).
For repeated failures, the CorrectionAgent compares the current error with earlier ones and explains in the correction prompt why the problem may be recurring, what mistake to avoid, and what fix to try~\cite{Wang_2024}.

\subsubsection{Correction strategies}

For each correction attempt, the pipeline supplies the correction agent with the current step-definition code, the error log from the failed iteraction, the execution plan, and the cumulative error history.
The correction agent supports three strategies, whose use is controlled by explicit rules and correction-specific prompts.
The pipeline determines which strategies to request based on the failure type, the state of the current file, and the configured correction policy.
Each strategy starts from the same version of the file.
The prompts instruct the model to preserve working code and prioritize the simplest correction that fixes the observed failure.

The agent first verifies the error type; if it is an infrastructure error, it retries execution with the current file.
Implementation errors fall into two fix alternatives: \textit{Full Regeneration} and \textit{Incremental Editing}.
The former is executed when the file is empty, the failure indicates truncated code, incremental editing is disabled, or the failure is classified as an infrastructure execution timeout.
The latter is triggered for failures routed to code correction when none of the above conditions applies.

The correction agent returns each proposed correction as either a SEARCH/REPLACE blocks or a JSON diff, contrasting mostly in the format the correction is expressed.
SEARCH/REPLACE copies the affected definition into the SEARCH block and supplies a REPLACE block with the corrected code, while JSON Diff returns a structured list of replacement, insertion, deletion, or append operations.
JSON Diff, in particular, returns the fix as a structured list of operations, enabling the pipeline to patch current file in specific locations.

The CorrectionAgent produces up to three correction suggestions and sends them to the Patch Reranker.
Each suggestion is labeled with the strategy that produced it, together with the starting file, the expected feature steps, and the error log.

\subsubsection{Patch reranker and cycle detection}

When \textit{CorrectionAgent} produces more than one possible fix, a patch reranker compares them and selects the highest-scoring patch.
This score is calculated based on five dimensions: (a) similarity level to the file's latest version; (b) relative edit distance (impact); (c) the proportion of scenario steps matched by step definitions, with penalties when multiple definitions match the same step; (d) the results of syntax, language-specific, and lint checks before Docker execution, and (e) relevance to the root error cause.

During the correction iteration, the Pipeline also monitors for the same error in consecutive attempts.
When it detects a cycle, a complete regeneration is triggered to prevent repeated unsuccessful corrections.
The final output of this stage is the highest-scoring valid patch applied to the current version of the generated file.

\subsection{RAG Subsystem}
\label{sec:rag}

The \textit{UnifiedStepsAgent} and \textit{CorrectionAgent} use BrinAI's RAG subsystem, \textit{RAG-GWT}, to retrieve examples that guide step-definition generation and repair, respectively.
These examples provide concrete references for implementing feature steps, helping the agents follow BDD conventions and reduce hallucinations~\cite{lewis2021retrievalaugmentedgenerationknowledgeintensivenlp}.
The knowledge base was derived from the GWT dataset~\cite{GivenWhenThenDataset}, where we removed all entries originating from the repositories included in GWT-Verified, leaving 2,183 entries.
Thus, the examples available for retrieval do not originate from any of the repositories used for verification, preventing repository overlap between RAG-GWT and our evaluation set.
We described in detail how RAG-GWT was built and how both \textit{UnifiedStepsAgent} and \textit{CorrectionAgent} retrieve the information stored in the RAG.

\noindentparagraph{Index construction.}

We structure each indexed chunk as a single step definition, aligning the scope level between the requested and generated source code.
Each chunk is indexed with a single dense embedding computed using Sentence-BERT (\texttt{all-MiniLM-L6-v2})~\cite{reimers2019sentencebertsentenceembeddingsusing}.
The embedding is derived from textual information, including the step pattern, keywords extracted from that pattern, and context from the source feature, allowing queries written in Gherkin to match step-definition chunks.
During the indexing procedure, we categorized the chunks into four application domains---UI, API, database, and CLI---to improve RAG suggestions.
For each chunk, we count specific keywords belonged to each domain---e.g., \texttt{click} and \texttt{visit} for UI, \texttt{status code} and \texttt{endpoint} for API---and assigned the chunk to the highest count, or to \textit{general} if all counts are zero.

\noindentparagraph{Retrieval and selection.}
Both agents interact with RAG differently, according to their specific needs.
For repair, the \textit{CorrectionAgent} combines the failed code and its error log into a query on the RAG-GWT index.
It retrieves individual step definitions in the target programming language based on cosine similarity.
For step-definition generation, the \textit{UnifiedStepsAgent} uses the complete target feature file as the query to retrieve semantically similar step definitions, filtered by programming language and domain.
Retrieved candidates are grouped and scored based on their average similarity, and up to two best-matching chunks are injected into the \textit{UnifiedStepsAgent}'s generation prompt.

\subsection{Memory Architecture}
\label{sec:memoria}

During our preliminary analysis, we observed that the agent failed to produce valid results when no project-specific conventions were provided.
We propose a memory architecture that supplements the agent's internal knowledge with explicit repository context, retrieved according to the task~\cite{zhang2025memory,Zhang2023RepoCoder}.
Following industry standards~\cite{huyen_ai_2025}, we define three memory layers with distinct scope and lifetime.

\begin{itemize}
  \item \textbf{Internal knowledge}
        corresponds to what is already encoded in the weights of the pre-trained model, such as general familiarity with Cucumber/BDD conventions, common libraries, and test idioms.
        This information is available without any additional infrastructure.
  \item \textbf{Short-term memory}
        encompasses the data produced and consumed during the execution of a single snapshot, being discarded afterwards.
        Among the information stored in this layer are the reasoning trajectory of each agent, the attempt history, and the accumulated error records.
  \item \textbf{Long-term memory}
        persists data beyond the scope of a single snapshot.
        A repository-based storage gathers the failures from all snapshots analyzed in the current session.
        When a subsequent snapshot comes from the same repository, the system retrieves those previous failures, summarize them into lessons learned, and injects it into the correction prompt.
\end{itemize}


\section{Evaluation}
\label{sec:evaluation}

\subsection{Research Questions}
\label{sec:research-questions}

We designed our evalution using the \textit{Goal-Question-Metric} (GQM) approach~\cite{basili1994gqm}.
This approach structures an evaluation around a goal, derives the research questions from that goal, and then leverage metrics to answer them.
\textbf{Our goal is to evaluate the \underline{cost} and \underline{effectiveness} of BrinAI at generating executable step definitions for BDD scenarios}.
To answer this goal, we defined the following research questions and present their respective metrics in Table~\ref{tab:rq-metricas}.

\begin{description}
  \item[\textbf{RQ1.}] \textbf{How effective is BrinAI?}
        Here, we define effectiveness as the ability of BrinAI to generate \textit{step definitions} that correctly implement BDD scenarios.
        Hence, we measure the number of snapshots with fully executed \textit{step definitions}, as well as the similarity level between the generated code and the reference implementations.
        We also performed an ablation study to understand the impact of our RAG subsystem at correctly generating these \textit{step definitions}.

  \item[\textbf{RQ2.}] \textbf{What is the cost of using BrinAI?}
        This question examines which and how much resources BrinAI requires to produce a solution.
        To answer this question, leveraged a telemetry mechanism to collect, for each snapshot, the number of tokens consumed, the number of attempts, and the number and types of tools used.
        We then calculated the cost of each execution and the cost per successful solution.

  \item[\textbf{RQ3.}] \textbf{To what extent does BrinAI outperform LLM baselines?}
        We hypothesize that agent's capabilities---such as planning, tool use, context selection, and iterative correction---provide a relevant advantage over a baseline that relies solely on LLMs.
        Thus, we designed an experiment to compare BrinAI with a single-shot LLM-baseline to generate the same \textit{step definitions} in terms of effectiveness and cost.
\end{description}


\begin{table}[htbp]
  \centering
  \caption{Metrics and analysis procedures associated with the research questions.}
  \label{tab:rq-metricas}
  \begin{tabularx}{\textwidth}{
      P{0.07\textwidth}
      P{0.18\textwidth}
      >{\RaggedRight\arraybackslash}X
    }
    \toprule
    \textbf{RQ}       &
    \textbf{Metric}   &
    \textbf{Definition}                                                                                                                                                       \\
    \midrule


    \rowcolor{gray!15}
    \textbf{RQ1}      &
    Success rate      &
    Proportion of snapshots successfully resolved, i.e.,~those that passed.                                                                                                   \\
    \textbf{RQ1}      &
    Execution Outcome &
    The outcome of each snapshot execution, including success, failure, and partial success.                                                                                  \\
    \rowcolor{gray!15}
    \textbf{RQ1}      &
    Similarity        &
    The similarity level of each \textit{step definition} generated by BrinAI with respect to the reference implementation, measured by the amount of method calls in common. \\
    \textbf{RQ1}      &
    RAG Impact        &
    A comparison between the effectiveness of BrinAI with and without our RAG subsystem.                                                                                      \\
    \midrule
    \rowcolor{gray!15}
    \textbf{RQ2}      &
    Average Cost      &
    Overall cost per execution, calculated as the total cost divided by the number of snapshots resolved.

    \\
    \textbf{RQ2}      &
    Tokens            &
    The amount of input and output tokens produced during the execution of each snapshot.                                                                                     \\
    \rowcolor{gray!15}
    \textbf{RQ2}      &
    Tool calls        &
    Number of tool calls made during the execution of each snapshot.                                                                                                          \\
    \textbf{RQ2}      &
    Tool use          &
    Number of distinct snapshots that used each tool at least once.
    Calls and snapshots per tool, as well as usage intensity.                                                                                                                 \\
    \midrule
    \rowcolor{gray!15}
    \textbf{RQ3}      &
    Success delta     &
    Comparison between the number of successful executions of BrinAI and each prompt in the baseline.                                                                         \\
    \textbf{RQ3}      &
    Similarity delta  &
    Comparison between the level of similarity produced by BrinAI and each prompt in the baseline.                                                                            \\
    \rowcolor{gray!15}
    \textbf{RQ3}      &
    Cost delta        &
    Cost difference, in terms of tokens consumption, between the execution of BrinAI and each prompt in the baseline.                                                         \\
    \bottomrule
  \end{tabularx}
\end{table}

\subsection{Leveraging our Baseline}
\label{sec:baseline}

As previously mentioned, we designed \textbf{RQ3} to compare an agent-based approach with a baseline.
As we did not identify prior research specifically addressing the generation of step definitions for BDD scenarios, we built a single-shot LLM-based baseline to serve as a default reference for this task.
We opted for this baseline because LLMs are widely used for code generation and have proven effective compared with previous state-of-the-art methods~\cite{Schafer2024, nijkamp2023codegen, xia2023repair}.

We executed the baseline using the same 171 snapshots from GWT-Light, providing the BDD scenario, the system code, and the target programming language as input to a single prompt.
The prompt instructed the LLM to generate a complete step definitions file.
We did not implement any improvement technique---such as re-prompting, repair, or refinement---to avoid adding agentic behavior to the baseline.

The remainder of this section details the design of the baseline, its experimental design, and the execution protocol.

\subsubsection{Selected Models}
\label{sec:baseline-models}

Our baseline was built upon two models: \textbf{openai/gpt-5-mini} and \textbf{minimax/minimax-m2.5}.
At the time of our experiments, these models were among the top performers in software engineering tasks, according to the SWE-bench leaderboard~\cite{jimenez2024swebench};
\textbf{MiniMax M2.5} featured in 6th and \textbf{GPT-5 Mini} in 13th place.\footnote{SWE-bench Leaderboard: \url{https://www.swebench.com/}. Accessed at sep. 3, 2026.}
The models gather distinct attributes in terms of structure and architecture.
\textbf{GPT-5 Mini} is a proprietary model developed by OpenAI, with a context window of 400,000 tokens and support for up to 128,000 output tokens.
As it is a proprietary model, its weights are not publicly available, and OpenAI does not disclose the number of parameters or details regarding its architecture.\footnote{\url{https://developers.openai.com/api/docs/models/gpt-5-mini}. Accessed at sep. 3, 2026.}
By contrast, the \textbf{MiniMax M2.5} model is an open-weight model based on a \textit{Mixture-of-Experts} (MoE) architecture, with approximately 230 billion total parameters.
The model contains a context window of approximately 200,000 tokens, and its weights are publicly available.\footnote{\url{https://huggingface.co/MiniMaxAI/MiniMax-M2.5}. Accessed at sep. 3, 2026.}
We used the models provided by OpenRouter, a well-known platform that provides access to multiple LLMs through a unified API, allowing us to easily switch between models and compare their performance in a single environment.\footnote{\url{https://openrouter.ai/}.}

\subsubsection{Prompt Strategies}
\label{sec:baseline-prompts}

We defined three prompt formats to build this baseline: \textbf{Zero-Shot}, \textbf{One-Shot}, and \textbf{Chain-of-Thought (CoT)}.
Given the nature of GWT-Light, we structured these strategies to dynamically incorporate not only the BDD scenario and the system code, but also general instructions about the programming language, the testing environment, and the expected output format.

\begin{figure}[htbp]
  \centering
  \includegraphics[width=0.8\textwidth]{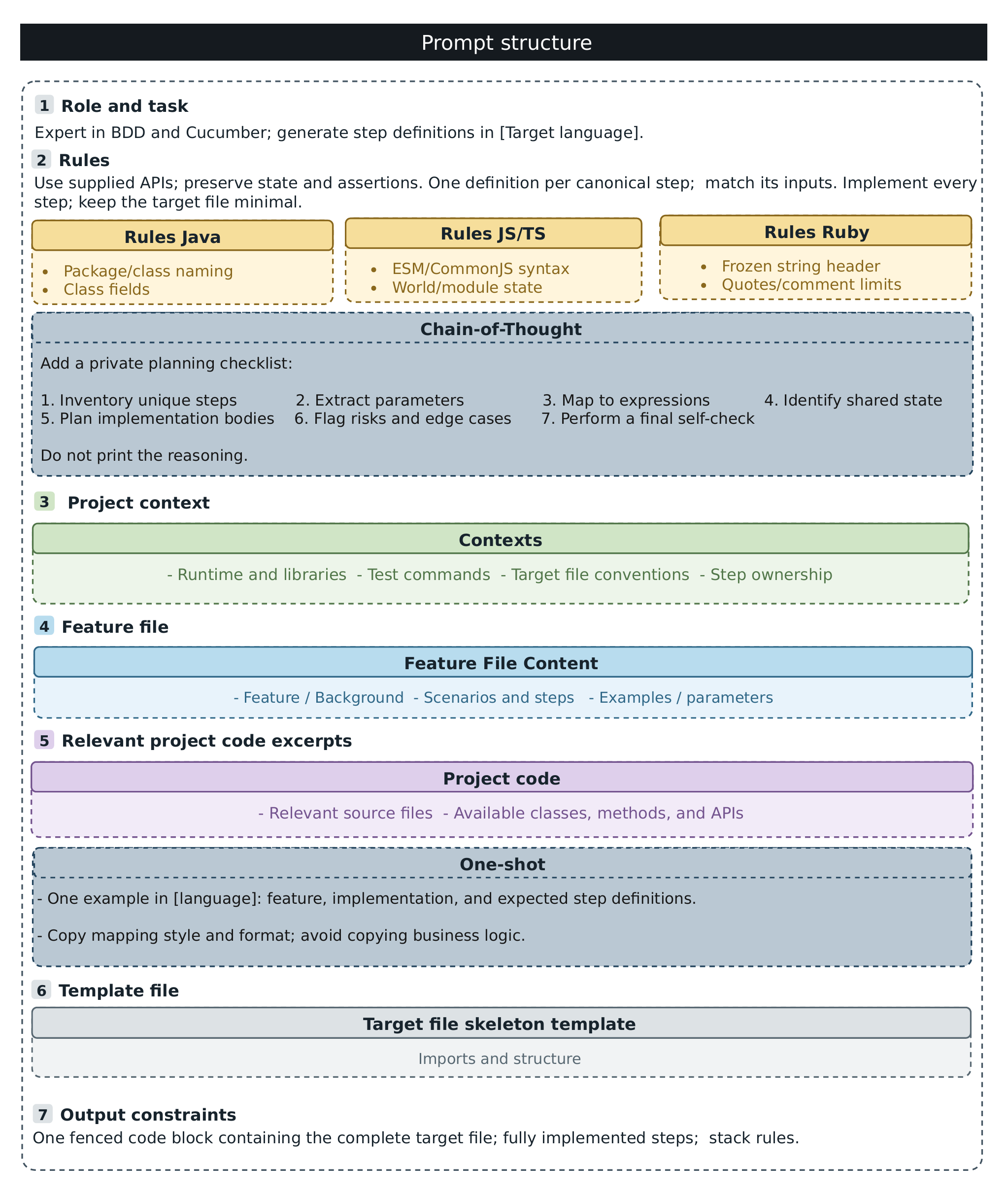}
  \caption{Default prompt structure shared across the three prompting strategies}
  \Description{Diagram illustrating the default prompt structure shared across the three prompting strategies. The prompt is organized into four main blocks, each grouping specific contextual information and instructions used to guide the model.}
  \label{fig:default-structure}
\end{figure}

\noindentparagraph{Default Structure.}
We followed best practices for prompt design and created default structure to be shared across all three strategies~\cite{amatriain2024promptdesignengineeringintroduction}.
As depicted in Figure~\ref{fig:default-structure}, this structure is organized in four major blocks.

\begin{enumerate}
  \item \textbf{Task and persona definition.} We start defining the overall goal for the task and the persona that the model should assume to perform it.
  \item \textbf{Rules.} We define a set of rules that the model must follow to generate the step definitions file.
        These rules are specific to the target programming language, and lists the allowed ecosystem, the expected imports and assertion libraries, how to define the step definitions, etc.
  \item \textbf{Project Context.} We provide the model with general information about the snapshot's project, including the technology stack, some dependencies and runtime tips (e.g.,~Java~17 with Selenium for browser tests, or Maven with REST Assured for API tests), the feature file to be implemented, and snippets of the system code that are relevant to the task.
  \item \textbf{Output structure.} This block defines the expected output format.
        We ask the model to produce a single code block containing the whole step definitions file, with all steps fully implemented and without any placeholders or empty bodies.
\end{enumerate}

\noindentparagraph{Zero-Shot.}
This prompt implements the default structure without any additional examples or specific instructions.
It represents the basic configuration of our baseline.

\noindentparagraph{One-Shot.}
This prompt extends the zero-shot strategy by including a complete, self-contained example of a BDD test scenario in \textit{Project Context}.
The example contains a feature file, an implementation file, and the corresponding step definitions for a hypothetical calculator we implemented for each programming language included in our evaluation (Java, Ruby, and JavaScript).

\noindentparagraph{Chain-of-Thought (CoT).}
Our third prompt strategy adds a seven-step structured reasoning checklist between the rules and the project context.
These steps guides the model to start leveraging all steps in the scenario, extract possible parameters, match them to the appropriate step expression, detect states to be shared between steps, plan which implementation to write and execute, and finally perform a sanity check to ensure that all steps are covered, without duplicates or empty bodies.

\subsubsection{Baseline Execution}
\label{sec:baseline-execution}

We submitted each GWT-Light snapshot to both models (Section~\ref{sec:baseline-models}) under all three prompt strategies (Section~\ref{sec:baseline-prompts}), configuring a total of $171 \times 2 \times 3 = \textbf{1,026}$ independent executions.
We configured both models with a temperature of 0.1 .
Each execution performs exactly one model call followed by the execution of the generated file against the target scenario.
We ensured that no information from one execution was reused in another;
every execution starts from scratch, without any memory from previous attempts.

\noindentparagraph{Metrics Collection.}
After each execution, we collected all metrics defined in Table~\ref{tab:rq-metricas} to answer \textbf{RQ3}.
From the test report produced by Cucumber, we extracted the number of steps that passed and failed, which allowed us to determine whether the execution was successful, partially successful, or failed.
We also collected the number of input and output tokens consumed by the model according to the API response.
In case the snapshot execution fails---i.e.,~does not produce a valid test report---we classified it into one of two failure categories: (a) \textbf{test generation failure}, when the model does not produce an extractable or complete output due to token limit exceeding, API timeout, etc; and (b) \textbf{test execution failure}, when a source code is generated but the execution of the test scenario fails before producing a valid report.

\subsection{BrinAI Execution}
\label{sec:brinai-execution}

To execute BrinAI, we relied on the GPT-5 mini model for all subagents, with a temperature set to 0.1.
We defined a maximum threshold of four attempts for each snapshot execution;
one default generation and up to three subsequent opportunities for the agent to correct or regenerate the code.
We restricted the number of attempts to four after a preliminary cost analysis on the execution over GWT-Light dataset to not exceed our budget.

We executed BrinAI against the GWT-Light dataset under two configurations: with the RAG subsystem enabled (\textit{RAG on}) and disabled (\textit{RAG off});
this setup allows us to evaluate the impact of RAG on the effectiveness of BrinAI.
Therefore, BrinAI was executed a total of 342 times ($171 \times 2$), once for each snapshot in both configurations.

\noindentparagraph{Metrics Collection.}
We collected all metrics defined in Table~\ref{tab:rq-metricas} to answer \textbf{RQ1}, \textbf{RQ2}, and \textbf{RQ3}.
Effectiveness metrics (\textbf{RQ1} \& \textbf{RQ3}) were collected from the test report produced by Cucumber.
We used the report generated from the first successful attempt for each snapshot, or the attempt with the highest number of passed steps in case of partial/non-passed.
Reproduction recall was calculated only for snapshots classified as \textit{Passed}, \textit{Partial}, or \textit{Non-passed}, excluding test generation and execution failures.

As for cost-based metrics (\textbf{RQ2} \& \textbf{RQ3}), we collected them from all attempts, including those that failed, to account for the total resources consumed by BrinAI in each execution.
We opted for this approach since BrinAI's final output takes into account all attempts, not only the successful ones.


\section{Results}
\label{sec:results}

\subsection{RQ1: How Effective is BrinAI?}
\label{sec:rq1}

\noindentparagraph{\bcircle{Overall results.}}

We first evaluate BrinAI in its default configuration, with RAG enabled.
We assign each run to one of five outcome based on its execution result.
\textit{Passed} denotes functional success, i.e.,~the generated step definitions execute the target BDD scenario with all steps passing. \textit{Partial} means that at least one step passes, but the run does not achieve functional success. \textit{Non-passed} refers to an executable scenario in which no steps pass.
Two categories distinguish failures before and during execution.
\textit{Test generation failure} covers missing or unextractable code, truncated model output, request timeouts, and provider unavailability, whereas \textit{test execution failure} covers failures attributed to the execution environment or test harness.

Overall, BrinAI achieved a \textit{Passed} outcome in \textbf{77 out of 171 cases (45.0\%)}.
Another 42 cases (24.6\%) achieved \textit{Partial} success, and 16 (9.4\%) were considered \textit{Non-passed}.
The remaining cases comprised 21 test generation failures (12.3\%) and 15 test execution failures (8.8\%), all of which were recorded as execution timeouts.

\begin{figure}[htbp]
  \centering
  \includegraphics[width=0.65\columnwidth]{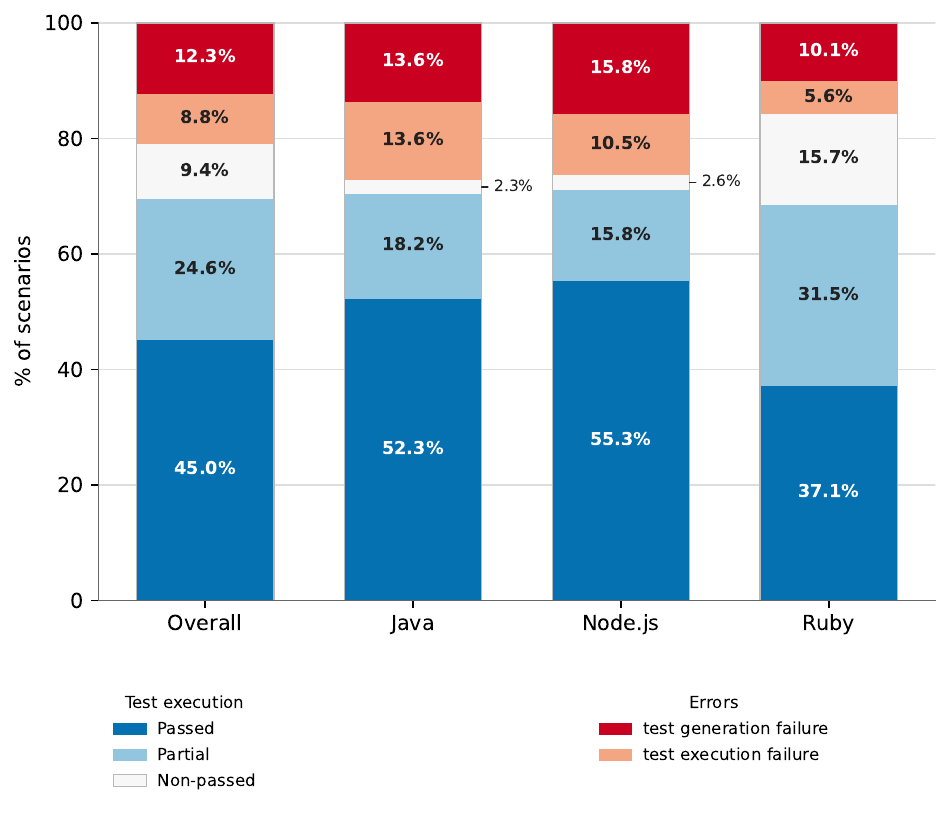}
  \caption{BrinAI effectiveness by programming language.}
  \Description{Bar chart showing the distribution of results from 171 runs of BrinAI with default configuration, categorized into complete success, partial success, model failure, and test execution failure.}
  \label{fig:rq1-overall}
\end{figure}

Figure~\ref{fig:rq1-overall} shows the distribution of these outcomes by programming language.
Our solution achieved the highest success rate in Node.js, where it solved \textbf{21 out of 38 scenarios (55.3\%)}, followed by Java, with \textbf{23 out of 44 (52.3\%)}; Ruby presented a lower success rate, achieving \textbf{33 out of 89 scenarios (37.1\%)}.
Nevertheless, BrinAI stand out when generating minimally executable scenarios in Ruby-based snapshots, as 84.4\% of them could be executed.

\noindentparagraph{\bcircle{Similarity Analysis.}}

Test execution success assesses whether the scenarios can be executed with the generated step definitions.
This is not sufficient to establish semantic correctness since the generated steps, although may be syntactically correct, can omit relevant application interactions or assertions.
This measure does not quantify by how much the generated implementation mirrors the expected behavior of the original one.

We address this gap with a new similarity-based metric, called \textit{reproduction recall}, where it measures the proportion of method calls in the reference implementation that are also present in the generated one.
We calculate reproduction recall at the method level, separately for each step definition; the reproduction recall of a given scenario is the average of the recalls for all its step definitions.
To compute this metric, we first detect the number of method calls in the reference implementation that are also present in the generated one (\textit{True Positive}, $TP$).
Then, we count the number of calls in the reference implementation that are not present in the generated one (\textit{False Negative}, $FN$).
Finally, we compute the \textit{reproduction recall} as $TP/(TP+FN)$.
We favored recall as we are interested in measuring how much of the original implementation is reproduced, rather than verifying whether the generated implementation exactly mirrors the original one.

\begin{table}[!ht]
  \centering
  \caption{Reproduction recall for each programming language.}
  \label{tab:rq1-recall}
  \begin{tabular}{ll|l|lllll}
    \toprule
    \textbf{Language} & \textbf{N} & \textbf{Average} & \textbf{Min.} & \textbf{Q1} & \textbf{Median} & \textbf{Q3} & \textbf{Max.} \\
    \midrule
    \textbf{Java}     & 32         & 0.561            & 0.000         & 0.390       & 0.549           & 0.750       & 1.000         \\
    \textbf{Node.js}  & 28         & 0.236            & 0.000         & 0.000       & 0.176           & 0.448       & 0.708         \\
    \textbf{Ruby}     & 75         & 0.378            & 0.000         & 0.223       & 0.333           & 0.516       & 1.000         \\
    \midrule
    \textbf{Overall}  & 135        & 0.392            & 0.000         & 0.200       & 0.350           & 0.573       & 1.000         \\
    \bottomrule
  \end{tabular}
  \par\smallskip
\end{table}

Table~\ref{tab:rq1-recall} presents the reproduction recall for all snapshots, grouped by programming language.
Overall, BrinAI achieved an average recall of \textbf{0.392} and a median of \textbf{0.350}.
In other words, at least half of the snapshots reproduced at most 35.0\% of the method calls implemented in the original step definitions.
BrinAI achieved the best performance in Java-based snapshots, with an average recall of \textbf{0.561} and a median of \textbf{0.549}.
Ruby features next with an average recall of \textbf{0.378} and a median of \textbf{0.333}.
Node.js snapshots gather the most different implementations in relation to their reference, with an average recall of \textbf{0.236} and a median of \textbf{0.176}, i.e.,~half of the snapshots reproduced at most 17.6\% of the method calls implemented in the original step definitions.

\begin{figure*}[t]
  \centering
  \includegraphics[width=\textwidth]{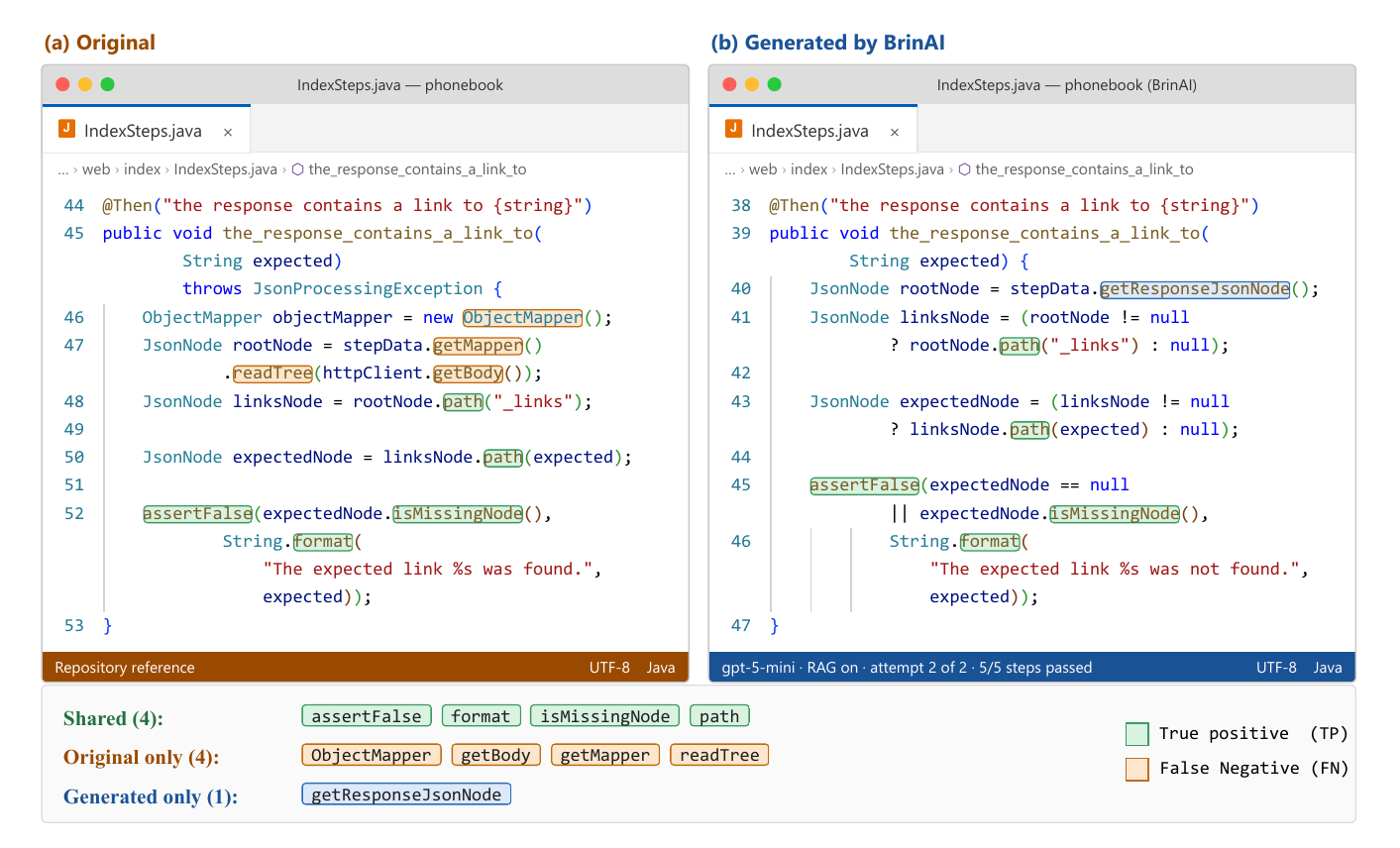}
  \caption{Excerpt from \texttt{IndexSteps.java} showing the reference and generated versions of the \texttt{Then} definition that checks for an expected link. Recall is 0.500 for the displayed definition and 0.8333 for whole the scenario.}
  \label{fig:recall-phonebook}
  \Description{Side-by-side comparison of the link-verification step definition. The reference parses the response body with readTree. BrinAI obtains the JSON tree with getResponseJsonNode and adds null checks. The versions share four of the eight distinct reference call names.}
\end{figure*}

\noindentparagraph{Similarity Examples.}

We report an analysis of two snapshots with contrasting recall values to illustrate how the generated step definitions differ from their reference implementations.

\paragraph{{Case \#1}}
Figure \ref{fig:recall-phonebook} shows an excerpt from \path{IndexSteps.java}, obtained from the project \path{Arc-E-Tect/arc-e-tect_phonebook}.\footnote{\url{https://github.com/Arc-E-Tect/arc-e-tect_phonebook}}
This snapshot corresponds to the scenario \textit{02 -- An API Consumer requests the top-level endpoints through the /index endpoint}, defined in \path{get_index.feature} file, and verifies if the index returns the expected links and relative paths.
BrinAI generated step definitions that passed all scenario steps with a recall of \textbf{0.8333}.

Both implementations share the logic for obtaining the \texttt{link} component of the response and verifying its presence, i.e.,~they first obtain \texttt{\_links} from the root node (\texttt{path("\_links")}), check whether they are in the expected path (\texttt{path(expected)}), and assert the link's presence (\texttt{isMissingNode()} and \texttt{format()}).
In fact, BrinAI improved the reference implementation by adding null checks every time it navigates the response tree, preventing null dereference errors during the test execution.
The difference between them lies in how they arrange the data to be checked.
While the reference implementation explicitly retrieves the \texttt{rootNode} from a new ObjectMapper, the generated implementation calls \texttt{getResponseJsonNode()}, a function that returns the stored HTTP response body as a \texttt{JsonNode}.

For the displayed \texttt{Then} step, four shared and four reference-only calls yield a \textit{reproduction recall} of \(4/(4+4)=0.500\) for that step.
Each distinct call name counts once, so the two calls to \texttt{path} contribute one match.
Although \texttt{getResponseJsonNode()} supplies the input needed for the check, this generated-only name does not affect recall.
The other two step definitions achieve full recall, yielding a \textit{reproduction call} of \((1+0.5+1)/3 \approx 0.8333\) for the scenario.

\begin{figure*}[t]
  \centering
  \includegraphics[width=\textwidth]{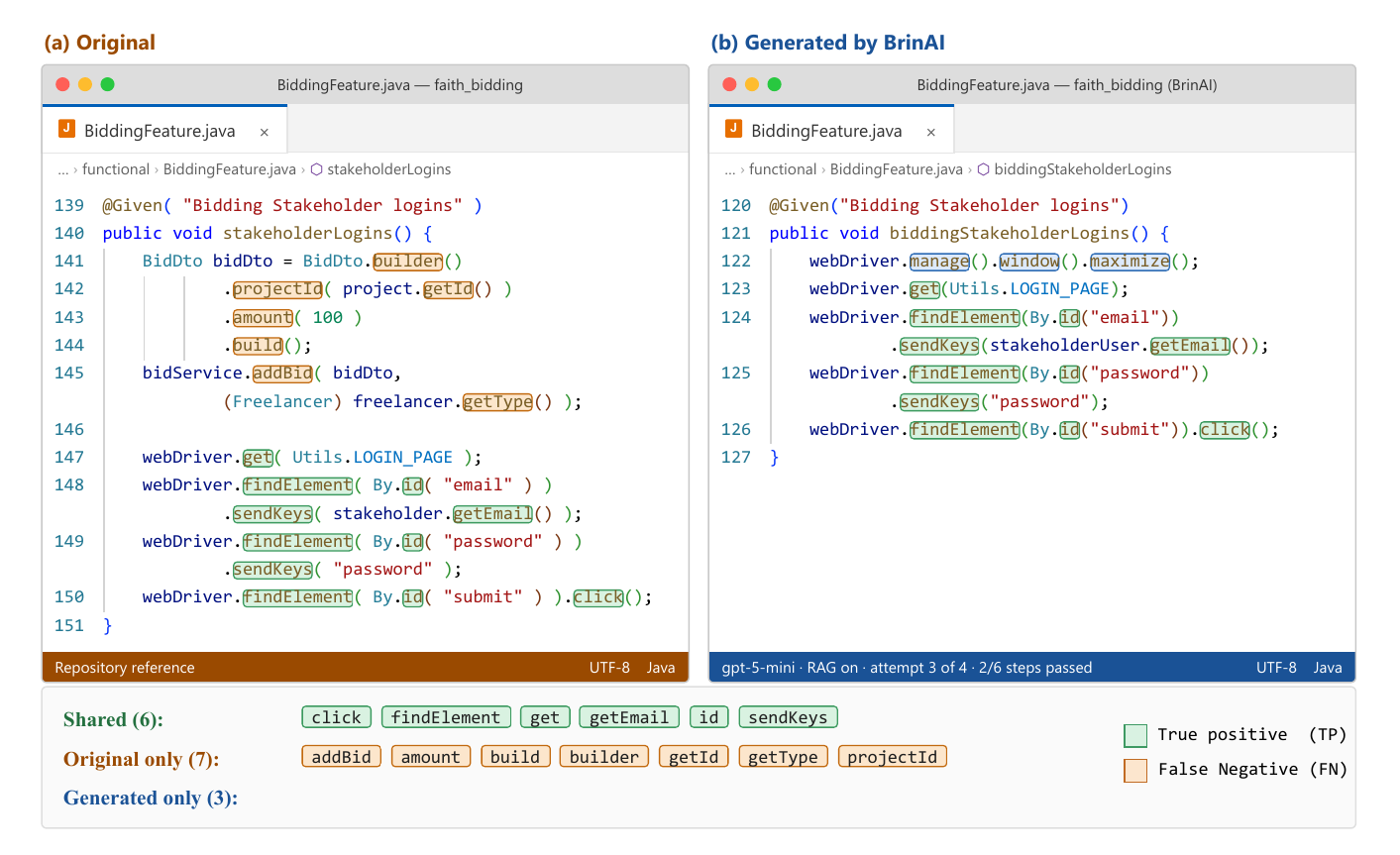}
  \caption{Excerpt from \texttt{BiddingFeature.java} showing the reference and generated stakeholder-login \texttt{Given}. Recall is 0.462 for the displayed definition and 0.436 for the scenario.}
  \label{fig:recall-faith}
  \Description{Side-by-side comparison of the stakeholder-login step. The reference creates a bid and then submits the login form. The generated step maximizes the browser window and submits the same form. Bid creation takes place in a Before hook outside the excerpt.}
\end{figure*}

\paragraph{{Case \#2}}
Figure~\ref{fig:recall-faith} shows an excerpt of \path{BiddingFeature.java}, obtained from the project \path{TheKiddos/faith}\footnote{\url{https://github.com/TheKiddos/faith}}.
The scenario \textit{Stakeholder Comment on Bidding}, defined in \path{Bidding.feature} file, is responsible for verifying the stakeholder's comment on a bidding is saved after clicking the add comment button.
The generated step definition achieved a partial success, with two of six steps passing; its reproduction recall was \textbf{0.436}.

Just like \textit{Case \#1}, both implementations share the method calls for acting on the login form.
The login page is opened with \texttt{get()} method, and all three form controls are located and interacted in the same way;
\texttt{findElement()} and \texttt{id()} locate them, \texttt{sendKeys()} enter the credentials, and \texttt{click()} submit the form.

Two major differences are observed in the beginning of the step definition.
The original implementation initializes \texttt{bidDto} using a builder (\texttt{builder()}), and sets its id (\texttt{projectId()}), and amount (\texttt{amount()}).
This object is then added to a bidService (\texttt{addBid()}).
The generated implementation does not include any of these calls, instead \underline{\textit{it implemented these steps in a new hook}}---\texttt{prepareData()}---executed before the login step;
interestingly, this hook sets the bid amount to 200 instead of 100.
The generated implementation also performed some specific actions, in this case to maximize the browser window (\texttt{manage().window().maximize()}) before opening the page.

For the displayed \texttt{Given}, the six calls match the reference implementation, while seven are considered missing, even though they are moved to the \texttt{prepareData()} hook.
This gives a \textit{reproduction recall} of \(6/(6+7) = 6/13 \approx 0.462\).
At the scenario level, the recall produced by each of the six step definitions are averaged.
The first fraction corresponds to the \texttt{Given} step shown in the figure, while the remaining five are computed as
\(
(6/13 + 3/3 + 3/8 + 4/6 + 1/9 + 0/1)/6
\approx 0.436.
\)

\noindentparagraph{\bcircle{RAG Ablation Study.}}

\begin{figure*}[t]
  \centering
  \includegraphics[width=\textwidth]{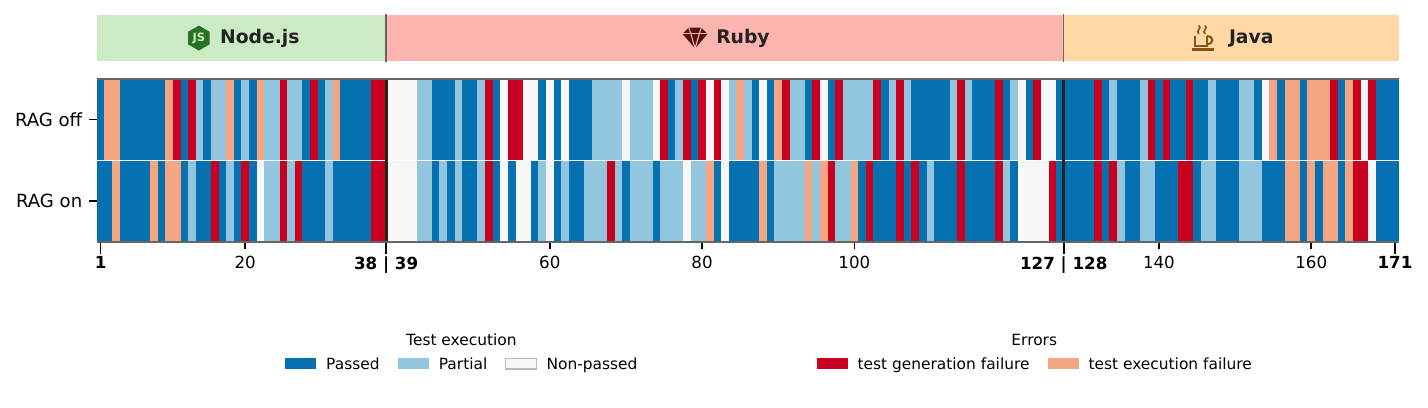}
  \caption{Comparison of BrinAI's effectiveness with RAG enabled and disabled.}
  \Description{Heatmaps comparing the results obtained by BrinAI for the same 171 scenarios under RAG-enabled and RAG-disabled configurations.}
  \label{fig:rq1-rag-ablation}
\end{figure*}

Finally, we conducted an ablation study to measure the impact of our RAG system at generating step definitions.
Figure~\ref{fig:rq1-rag-ablation} depicts a heatmap comparing the effectiveness status of BrinAI with RAG enabled (\textit{RAG on}) and disabled (\textit{RAG off}).
In this visualization, columns depict each of the 171 snapshots, while rows represent both configurations.
The color of each cell indicates the effectiveness level of BrinAI for that snapshot: success, partial success, no success, test execution failure, or test generation failure.
Also, we organized the snapshots by programming language, with Node.js on the left, followed by Ruby and Java.

As we can see, the impact of our RAG system is limited.
Both configurations achieved the same effectiveness result for 102 snapshots (59.6\%), while \textit{RAG on} and \textit{RAG off} performed better in 41 and 28 cases (24.0\%, 16.4\%), respectively.
Disabling RAG reduced the number of successful snapshots from 77 to 69 ($\Delta = -4.7\%$);
similar differences are observed for partial success (42 to 39, $\Delta = -1.8\%$) and no success (16 to 20, $\Delta = +2.3\%$).

We applied Wilcoxon signed-rank test on both \textit{RAG on} and \textit{RAG off} results to determine whether the differences are statistically significant~\cite{arcuri2014hitchhiker};
the test did not provide evidence of a systematic difference between them ($p = 0.21$).
Even though, we noticed that the inclusion of RAG flipped the effectiveness status from \textit{failure} to \textit{executable} in 24 snapshots ($\Delta = +14.0\%$).
That is, despite the limited overall impact, RAG assisted BrinAI in generating more functionally correct step definitions.

\begin{rqsummary}{RQ1 Summary}
  BrinAI was able to generate fully executable step definitions for 77 out of 171 snapshots (45.0\%), with another 42 partially executed (24.6\%).
  Its effectiveness varies across programming languages;

  Node.js-based snapshots achieved the highest success rate (55.3\%), while Ruby-based ones presented the lowest (37.1\%).
  Despite this, BrinAI generated implementations are only moderately similar to the original ones, i.e.,~its average and median reproduction recall stayed at 0.392 and 0.350, respectively.
  Finally, our RAG system contributed to producing more functionally correct step definitions.
\end{rqsummary}



\subsection{RQ2: What is the cost of using BrinAI?}
\label{sec:rq2}

\noindentparagraph{\bcircle{Cost Analysis.}}
Table \ref{tab:rq2-cost-results} lists the cost-related metrics collected during the execution of BrinAI on the GWT-Light dataset.
For each metric, we report the mean, median, first and third quartiles.
On average, each execution spent \textbf{US\$0.0973}, totaling \textbf{US\$16.64} across all executions.
The median cost was \textbf{US\$0.0844}, with most executions costing between \textbf{US\$0.0387} (Q1) and \textbf{US\$0.1378} (Q3).
The most expensive execution cost \textbf{US\$0.3568}, while the cheapest one spent \textbf{US\$0.0096}.
When we analyze token consumption, BrinAI consumed an average of \textbf{426,945 input tokens} and \textbf{33,530 output tokens}; an input-to-output ratio of \textbf{12.7}.
In fact, this ratio consistently increases throughout the token consumption spectrum;
while Q1 presents a ratio of \textbf{10.1} (122,688 and 12,141.5), Q2 and Q3 present ratios of \textbf{16.7} (336,830 and 20,131) and \textbf{17.1} (544,394 and 31,795.5), respectively.
Across the 171 executions, input token count was positively correlated with total execution cost (Pearson's $r = 0.51$).

\begin{table}[!t]
  \centering
  \caption{The amount of tokens consumed and the financial cost of executing BrinAI on the GWT-Light dataset.}
  \label{tab:rq2-cost-results}
  \begin{tabular}{l|l|lllll}
    \toprule
                                   & \textbf{Average} & \textbf{Min.} & \textbf{Q1} & \textbf{Median} & \textbf{Q3} & \textbf{Max.} \\
    \midrule
    \textbf{Financial Cost (US\$)} & 0.097284         & 0.009559      & 0.038690    & 0.084430        & 0.137788    & 0.356809      \\
    \textbf{Input Tokens}          & 426,945.02       & 17,866        & 122,688     & 336,830         & 544,394     & 2,732,941     \\
    \textbf{Output Tokens}         & 33,530.43        & 3,622         & 12,141.5    & 20,131          & 31,795.5    & 249,494       \\
    \bottomrule
  \end{tabular}
  \par\smallskip
\end{table}

\begin{figure*}[ht]
  \centering
  \includegraphics[width=.85\textwidth]{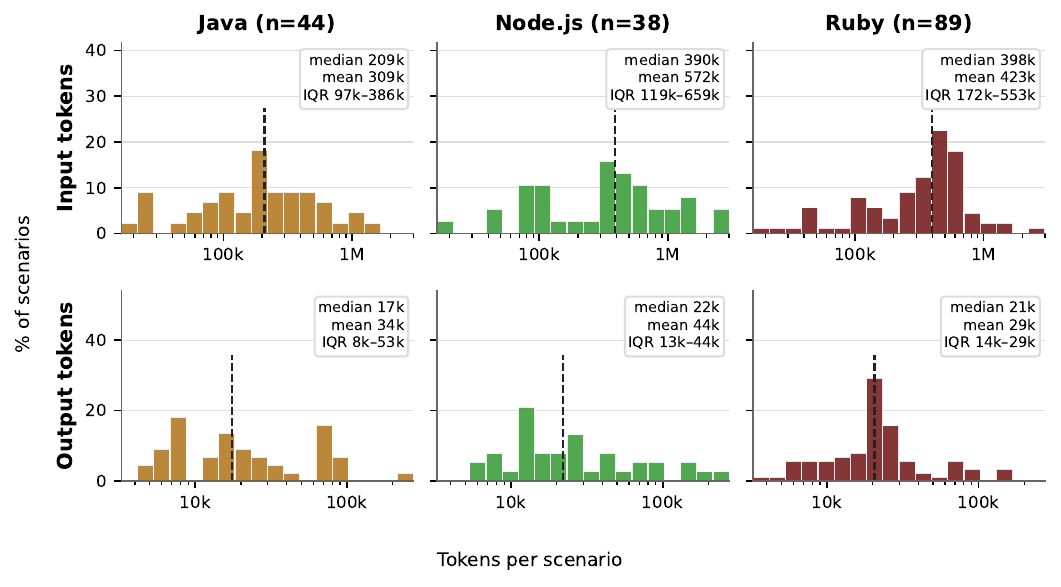}
  \caption{Distribution of input (top) and output (bottom) tokens per scenario
    by programming language. Bins are logarithmic and shared within each row; the
    y-axis shows the percentage of the language's scenarios. The dashed line marks
    the median.}
  \Description{}
  \label{fig:token-consumption-histogram}
\end{figure*}

Figure~\ref{fig:token-consumption-histogram} presents the distribution of input and output tokens across programming languages.
Input token consumption exhibits a right tail across all languages, driven by a few executions with high token consumption.
This effect is more pronounced in Java and Node.js, whose means are approximately 1.5 times their medians, while Ruby presents a mean close to its median.
Java presents the lowest median (208,917), followed by Node.js (389,668) and Ruby (397,926).
Node.js also presents the highest proportion of executions exceeding one million input tokens (18.4\%), compared to Java (4.5\%) and Ruby (4.5\%).
In contrast, output token consumption is more similar across languages, with medians of 17,423 (Java), 21,916 (Node.js), and 20,592 (Ruby).
Overall, these results indicate that input token consumption exhibits greater variability across programming languages, particularly in Node.js executions.

\noindentparagraph{\bcircle{Tool Calls.}}
One of the major advantages of a multi-agent system is the ability to dynamically interact with external tools to retrieve information and perform actions~\cite{huyen_ai_2025}.
We analyzed the number of tool calls made by BrinAI during its execution on each snapshot.
On average, BrinAI invoked external tools \textbf{19.39 times per snapshot}, with a median of 17 calls.
Most snapshots required between 12 and 25 calls (Q1 and Q3, respectively), while the minimum and maximum number of calls were 3 and 68.

\begin{figure*}[htbp]
  \centering
  \includegraphics[width=0.85\textwidth]{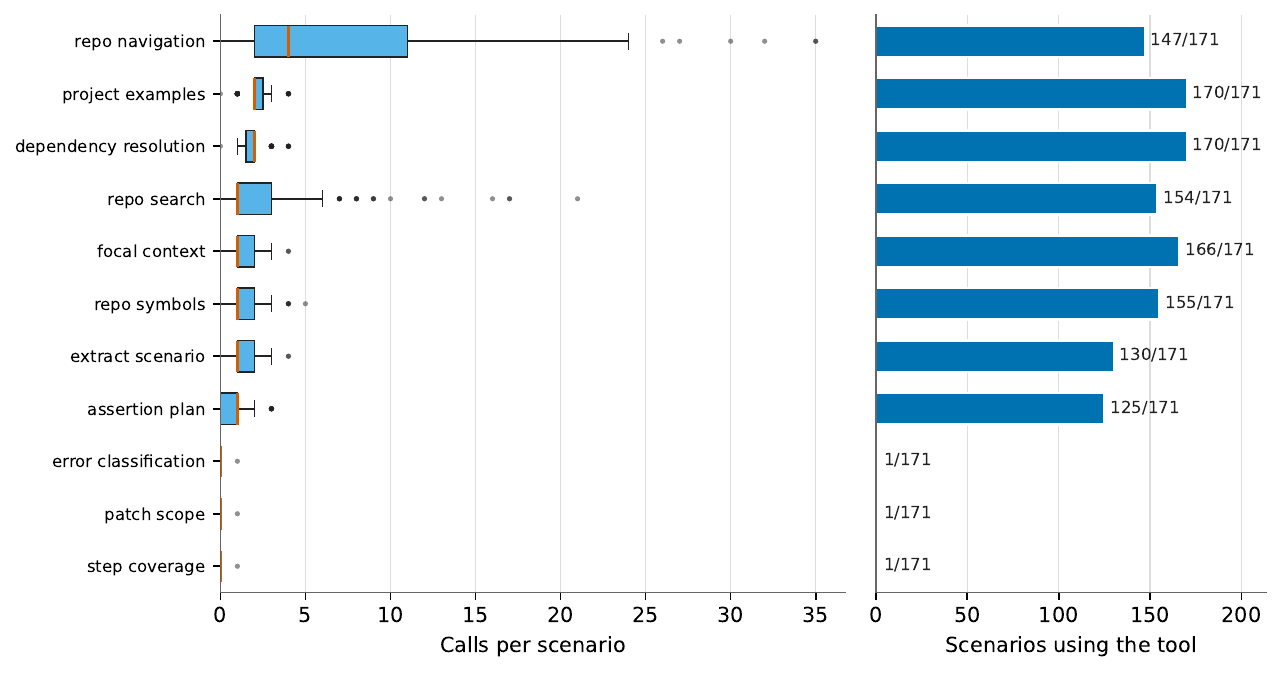}
  \caption{Distribution of tool calls performed by BrinAI, broken down by tool type.}
  \Description{Distribution of the total number of tool calls per scenario and the number of calls for each tool used by BrinAI.}
  \label{fig:rq2-tools}
\end{figure*}

Figure~\ref{fig:rq2-tools} shows the amount and distribution of calls of each tool across all snapshots.
Most tools are frequently invoked by BrinAI, i.e.,~eight out of the 11 tools were called in at least 125 snapshots (73\%).
Two stand out as the most frequently used: \textit{project examples} and \textit{dependency resolution}, invoked in \textbf{170 out of 171 snapshots}, each;
\textit{focal context} tool follows next being called in 166 snapshots.
These tools extract reusable patterns from existing step definitions, recommend HTTP clients and assertion libraries based on project dependencies, and retrieve scenario-relevant code snippets, routes, and identifiers, respectively.
Repository-related tools---such as \textit{repo symbols}, \textit{repo search}, and \textit{repo navigation}---were called in 155, 154, and 147 snapshots, respectively.
On the opposite side, \textit{step coverage}, \textit{patch scope}, and \textit{error classification} tools were used in one snapshot, each.

When we analyze the number of calls per scenario, \textit{repo navigation} stands out as the most frequently called;
the agent invoked this tool \textbf{4} times per snapshot on median (Q1 = 2, Q3 = 11).
Both \textit{dependency resolution} and \textit{project examples} were called \textbf{2} times per snapshot on median (Q1 = 1.5, Q3 = 2).
Other frequently called tools---e.g.,~\textit{repo search}, \textit{focal context}, \textit{extract scenario}, etc---were called once per snapshot on median.

\begin{rqsummary}{RQ2 Summary}
  The average cost of executing BrinAI on a snapshot is \textbf{US\$0.0973}, with a median of \textbf{US\$0.0844}.
  Input and output token consumption averaged \textbf{426,945} and \textbf{33,530}, respectively;
  an input-to-output ratio of \textbf{12.7}.
  BrinAI frequently invoked external tools, with a median of 17 calls per snapshot;
  most snapshots required between 12 and 25 calls (Q1 and Q3, respectively), while the minimum and maximum number of calls were 3 and 68.
  Although \textit{project examples} and \textit{dependency resolution} are virtually called in every snapshot (170 out of 171), \textit{repo navigation} is the most frequently invoked one, with a median of 4 calls per snapshot.
\end{rqsummary}

%

\subsection{RQ3: To what extent does BrinAI outperform the LLM baselines?}
\label{sec:rq3}


As stated previously in Section~\ref{sec:baseline-models}, we compared BrinAI with three prompt engineering strategies: zero-shot, one-shot, and Chain-of-Thought (CoT).
We report this comparison in terms of effectiveness, reproduction recall, and cost.

\begin{figure*}[ht]
  \centering
  \includegraphics[width=0.75\textwidth]{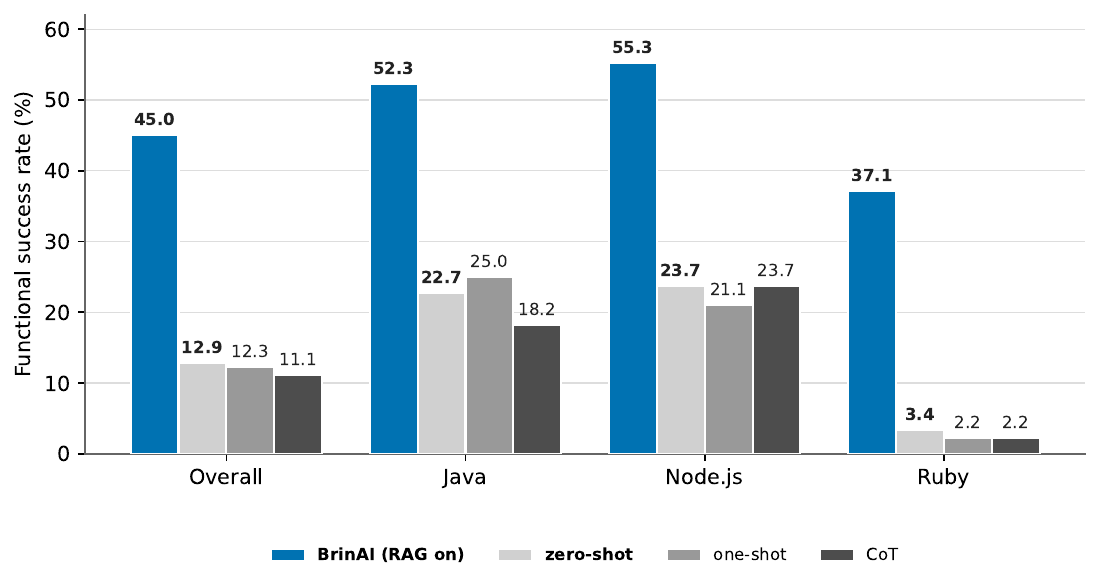}
  \caption{Full success rate of BrinAI compared with the three prompt engineering strategies: zero-shot, one-shot, and Chain-of-Thought.}
  \Description{Gráfico comparando as taxas de sucesso funcional do BrinAI e dos baselines zero-shot, one-shot e Chain-of-Thought.}
  \label{fig:rq3-effectiveness}
\end{figure*}

\noindentparagraph{\bcircle{Effectiveness Comparison.}}
Figure~\ref{fig:rq3-effectiveness} compares the effectiveness of BrinAI with the three prompt engineering strategies on the 171 snapshots evaluated, where we report the number of fully executable snapshots for each approach.
While BrinAI solved 45\% of the snapshots, the prompt-based approaches solved 12.9\% of the cases at most (zero-shot), an absolute difference of 32.1 p.p.
This difference holds across programming languages, with BrinAI outperforming the baselines by 27.3, 31.6, and 33.7 p.p. in Java, Node.js, and Ruby, respectively.

\begin{figure*}[ht]
  \centering
  \includegraphics[width=0.85\textwidth]{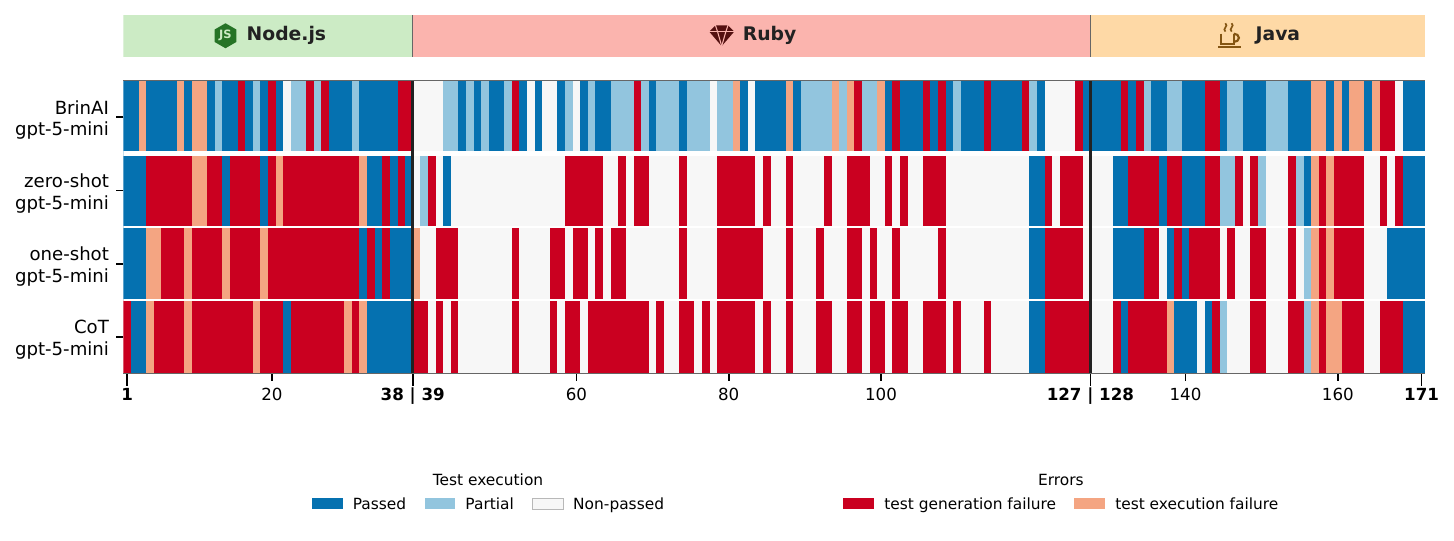}
  \caption{Individual results by scenario for BrinAI and for the zero-shot, one-shot, and Chain-of-Thought prompt engineering strategies.}
  \Description{Heatmap showing, for each of the 171 scenarios evaluated, the success or failure of BrinAI and the three prompt engineering strategies: zero-shot, one-shot, and Chain-of-Thought.}
  \label{fig:rq3-scenarios}
\end{figure*}

\noindentparagraph{Individual Scenario Analysis.}
Figure \ref{fig:rq3-scenarios} depicts the effectiveness heatmap of BrinAI and all three prompt strategies per snapshot.
At first glance, we observe a significant higher presence of failures when compared to BrinAI.
This is especially evident for Node.js, where 10 out of the 38 (26.3\%) snapshots presented some failure in BrinAI, while 29 (76.3\%), 30 (78.9\%), and 29 (76.3\%) snapshots failed in zero-shot, one-shot, and CoT, respectively.
Despite the lower number of failures in Ruby and Java-based snapshots, we noticed a prevalence of snapshots that were executed but failed at the first step, represented by the light gray color in the heatmap.
Put differently, prompt engineering strategies frequently do not generate a step definition as a result;
whey they do, the generated step is often non functional, leading to a failed execution.
Nevertheless, there were cases where LLM-based approaches outperformed BrinAI: zero-shot did so in 14 out of 171 (8.2\%) snapshots, one-shot in 17 (9.9\%), and CoT in 11 (6.4\%).

\noindentparagraph{\bcircle{Similarity Analysis.}}
As zero-shot presented the highest success rate among the baseline strategies, we computed the \textit{reproduction recall} for this strategy, and compared it with BrinAI.
We included only snapshots classified as \textit{Passed}, \textit{Partial}, or \textit{Non-passed} with a recorded recall value, retaining zeros and excluding test generation and execution failures.
This resulted in 135 eligible snapshots for BrinAI and 92 for zero-shot.
This comparison is summarized in Table~\ref{tab:zero-shot-similarity}.

\begin{table}[ht]
  \centering
  \caption{Similarity between BrinAI and Zero-Shot in terms of reproduction recall.}
  \label{tab:zero-shot-similarity}
  \begin{tabular}{ll|l|lllll}
    \toprule
                                        & \textbf{Language} & \textbf{Average} & \textbf{Min} & \textbf{Q1} & \textbf{Median} & \textbf{Q3} & \textbf{Max} \\
    \midrule
    \multirow{3}{*}{\textbf{BrinAI}}    & \textbf{Java}     & 0.561            & 0.000        & 0.390       & 0.549           & 0.750       & 1.000        \\
                                        & \textbf{Node.js}  & 0.236            & 0.000        & 0.000       & 0.176           & 0.448       & 0.708        \\
                                        & \textbf{Ruby}     & 0.378            & 0.000        & 0.223       & 0.333           & 0.516       & 1.000        \\
    \midrule
    \multirow{3}{*}{\textbf{Zero-shot}} & \textbf{Java}     & 0.343            & 0.000        & 0.060       & 0.325           & 0.547       & 1.000        \\
                                        & \textbf{Node.js}  & 0.383            & 0.000        & 0.200       & 0.333           & 0.333       & 1.000        \\
                                        & \textbf{Ruby}     & 0.232            & 0.000        & 0.000       & 0.200           & 0.333       & 1.000        \\
    \bottomrule
  \end{tabular}
  \par\smallskip
\end{table}

BrinAI achieved higher median reproduction recall than zero-shot for Java and Ruby.
The largest median difference was observed for Java (0.549 for BrinAI versus 0.325 for zero-shot), followed by Ruby (0.333 versus 0.200).
Node.js was the only language for which zero-shot achieved higher median recall (0.333 for zero-shot versus 0.176 for BrinAI).

\noindentparagraph{\bcircle{Cost Difference.}}
Table~\ref{tab:rq3-zero-shot-costs} reports the cost for applying zero-shot to the 171 snapshots; BrinAI's corresponding results are shown in Table~\ref{tab:rq2-cost-results}.
Although BrinAI achieves a higher success rate, this improvement comes at greater expense.
On average, BrinAI requires 0.0973 USD per snapshot, compared with 0.0053 USD for zero-shot.
The average cost per successful solution is 0.2160 USD for BrinAI and 0.0408 USD for zero-shot.

\begin{table}[!ht]
  \centering
  \caption{The amount of tokens consumed and the financial cost after executing Zero-shot.}
  \label{tab:rq3-zero-shot-costs}
  \begin{tabular}{l|l|lllll}
    \toprule
                                   & \textbf{Average} & \textbf{Min} & \textbf{Q1} & \textbf{Median} & \textbf{Q3} & \textbf{Max} \\
    \midrule
    \textbf{Financial Cost (US\$)} & 0.005256         & 0.001485     & 0.004287    & 0.006002        & 0.006223    & 0.007108     \\
    \textbf{Input Tokens}          & 2,336.94         & 819          & 1,755.5     & 2,171           & 2,712.5     & 6,143        \\
    \textbf{Output Tokens}         & 1,916.39         & 0            & 1,192.5     & 2,269           & 2,965.5     & 3,000        \\
    \bottomrule
  \end{tabular}
  \par\smallskip
\end{table}

In terms of token consumption, BrinAI used an average of 426,945 input tokens per snapshot, approximately 183 higher than zero-shot (2,337).
This difference was smaller for output tokens, with BrinAI averaging 33,530 per snapshot compared with 1,916 for zero-shot, a ratio of 17.5:1.
Even the lowest input and output token counts for BrinAI (17,866 and 3,622, respectively) exceeded the corresponding maxima for zero-shot (6,143 and 3,000).

\begin{rqsummary}{RQ3 Summary}
  BrinAI outperforms all three prompt strategies.
  While its success rate was \textbf{45.0\%}, prompt engineering strategies achieved between \textbf{11.1\%} and \textbf{12.9\%}; an absolute difference of \textbf{32.1 p.p}.
  Prompt-based solutions regularly fail to produce minimal executable step definitions.
  For example, while BrinAI failed to execute 10 (26\%) Node.js-based snapshots, zero-shot, one-shot, and CoT failed in 29 (76\%), 30 (79\%), and 29 (76\%), respectively.
  BrinAI also achieved higher average reproduction recall than zero-shot in Java and Ruby, although zero-shot performed better in Node.js.
  BrinAI's higher effectiveness comes at a substantially greater cost, averaging \textbf{US\$0.0973} per snapshot compared to \textbf{US\$0.0053} for zero-shot, approximately \textbf{18.5 times higher}.
\end{rqsummary}


\section{Discussion}
\label{sec:discussion}

The results obtained in our experiments have revealed important lessons about the use of agent-based approaches for automating the execution of BDD scenarios.
In this section, we discuss the main findings and their implications.

\noindentparagraph{\textbf{BDD automation requires more than generating syntactically plausible code.}}
We observe a significant gap between the success rate of BrinAI and its reproduction recall, i.e.,~45.0\% of fully executable snapshots versus 0.392 reproduction recall on average (Section~\ref{sec:rq1}).
This indicates that, differently from previous approaches ~\cite{Schafer2024, trabalho_altino_ast}, verifying the effectiveness of automatically generated step definitions for a BDD scenario goes beyond executing it.
In our view, the high abstract and scope levels of BDD scenarios---often written at the acceptance level---add another layer of difficulty to the task, as a deeper understanding of the project context and its domain are required.
Therefore, a two-fold evaluation is highly recommended for this kind of task, one focused on functional correctness and another on reproduction of the original implementation.

\noindentparagraph{\textbf{Agent-based architecture favors BDD automation.}}
As discussed in Section~\ref{sec:rq3}, BrinAI outperforms the best prompt engineering strategy (zero-shot) by 32.1 p.p. in terms of success rate (45.0\% vs 12.9\%).
Considering the 149 snapshots in which zero-shot did not achieve functional success, BrinAI achieved a better outcome in 119 (79.9\%).
Among those, 60 achieved full functional success, and 59 improved to an intermediate executable state, i.e.,~partial success or execution with no passed steps.
The remaining 30 snapshots stayed at the same outcome level (27) or regressed (3).
Altogether, these results suggest that the iterative nature of an agent-based approach---planning, context exploration, execution, and correction---offers useful capabilities for this task.
This hypothesis have been discussed in the literature in other contexts ~\cite{xia2024agentlessdemystifyingllmbasedsoftware, yang2024sweagentagentcomputerinterfacesenable, zhang2024autocoderoverautonomousprogramimprovement, antoniades2025swesearch};
our results provide empirical evidence for BDD automation as well.

By contrast, our RAG system did not show a significant impact on the overall effectiveness of BrinAI.
Even though RAG increased the success rate from 40.4\% to 45.0\% ($\Delta = +4.6$ p.p.), there was no statistically difference between both configurations (see Section~\ref{sec:rq1}, RAG Ablation Study).
Similarly, Jain et al.~\cite{jain2024mitigatingcodellmhallucinations} conducted a RAG ablation study for API usage.
Their results showed that retrieval improves performance for low-frequency APIs but can degrade it for high-frequency APIs when irrelevant documentation is retrieved.
These findings highlight that RAG technique must be structured accordingly to achieve consistent gains, representing a particular challenge for this kind of system~\cite{kastner}.
Thus, further research is needed to find better ways to leverage RAG in agent-based approaches for BDD automation.

\noindentparagraph{\textbf{We must be aware of the cost of agent-based approaches.}}
If BrinAI clearly outperforms prompt engineering strategies, it does so at a higher cost.
As shown in Section~\ref{sec:rq2}, our agent-based approach consumed an average of 460,475 tokens, with 100 out of 171 snapshots (58.48\%) executing the whole four attempts before reaching a solution.
Furthermore, BrinAI performed an average of 19.39 tool calls per snapshot, with most of them related to repository exploration and semantic search.
Practitioners considering the use of an agent-based approach for BDD automation should think on strategies to reduce the cost of execution.
For example, configuration files---such as~\texttt{AGENT.md}, \texttt{SKILLS.md}, etc---can provide additional information about the project structure and its domain to the agent, potentially reducing its need for exploration and context retrieval ~\cite{helio_agents, helio_scam, andre_msr}.

\noindentparagraph{\textbf{Language ecosystem makes a difference.}}
BrinAI's effectiveness greatly varies between Ruby (37.1\%), Node.js (55.3\%) and Java (52.3\%) ecosystems.
Its hard to pinpoint any reason for this difference based on our evaluation, as Cucumber is a mature framework and shares the same structure, principles, and syntax across languages.
Previous experiments with LLMs have shown that the quality of generated code can vary across programming languages ~\cite{improta2025qualityinqualityout,sun2026evaluatinglanguagemodelscrosslanguage,macedo2024intertransleveragingtransitiveintermediate}.
Future research might uncover the reasons for such differences, and analyze some strategies to mitigate them.


\section{Threats to Validity}
\label{sec:threats}

We acknowledge that our study has some limitations that may affect the extension of our findings.
We discuss these main threats in this section.

\noindentparagraph{Construct Validity.}
BrinAI effectiveness is primarily evaluated based on its functional success, i.e.,~how many scenarios have been executed and passed after their steps were generated.
Although this measure shows that the generated step definitions are syntactically correct and functional, it does not show whether they test the same behavior as the original implementation.
We argue that a generated step definition approximates the original implementation if it performs the same method calls.
From this perspective, we leveraged a new metric---\textit{reproduction recall}---to measure how many method calls from the original implementation are present in the generated step definitions.
Still, same verification can be achieved by different implementations, and a low reproduction recall does not necessarily indicate a wrong solution.

\noindentparagraph{Internal Validity.}
We set up BrinAI to use GPT-5 mini with a temperature of 0.1, and adopted a single execution strategy to evaluate its effectiveness and cost.
Although we restrict our analysis to GPT-5 mini, this model was selected based on
(a) its performance in SWEBench~\cite{jimenez2024swebench}, a benchmark for evaluating LLMs in software engineering tasks, and
(b) its consistent superiority over MiniMax M2.5, another well ranked LLM in SWEBench.
Nevertheless, our results should be interpreted strictly in the context of our experimental setup, i.e.,~one specific model and configuration.

In our experiment, we intentionally simplified the execution strategy for LLM-only approaches to a single prompt-based generation, while previous work has shown that iterative prompting can improve the quality of the generated solutions~\cite{Schafer2024, xia2024agentlessdemystifyingllmbasedsoftware}.
While this design decision certainly limited baseline's performance, our goal was to compare the effectiveness of an agent-based approach---which can plan, execute, and iteratively correct its solutions~\cite{huyen_ai_2025}---against others relying on prompt engineering only.
Either way, our results do not allow us to attribute which BrinAI component is responsible for its superiority;
additional studies should be conducted to properly detect these contributions.

Finally, in order to reduce eventual data contamination between the execution of our scenarios, we used fully independent and previously validated snapshots.
We also ensured our RAG system was built upon different repositories than those used in evaluation, making it oblivious to the source code of the scenarios under generation.

\noindentparagraph{External Validity.}
We conducted our evaluation using 171 out of 713 GWT-Verified snapshots; a subset we called GWT-Light.
Although GWT-Light is 24\% of the size, these scenarios covers 96.3\% of the structural diversity reported in GWT-Verified.
Such selection was possible due to the stratified sampling process we applied~\cite{Nagappan2013Diversity}.
Yet, this does not mean our results can be generalized to BDD scenarios in general, as GWT-Verified was initially built from open-source projects and contains just a fraction of the programming languages supported by Cucumber.


\section{Related Work}
\label{sec:related-work}

\noindentparagraph{BDD Practices and Artifact Maintenance.}
Prior research focused mostly on understanding BDD adoption, its benefits, and the maintenance of its artifacts.
In a survey of 75 practitioners across 26 countries, \citet{Binamungu2018} found that BDD improves communication and provides executable specifications, but these artifacts become costly to maintain as they grow.
\citet{Zampetti2020} studied BDD adoption in 50,000 open-source projects, artifact co-evolution in 20 Ruby projects, and feedback from 31 developers.
They found that scenarios and fixtures often changed alongside or after production code, and that BDD frameworks were frequently used for other purposes, such as unit testing.
Their findings indicate that framework adoption alone does not imply following BDD practices.
In a systematic mapping of 166 papers, \citet{binamungu2023bdd} identified improvements to BDD processes and artifacts as recurring research topics, alongside a shortage of metrics and industrial evidence.

\noindentparagraph{LLM-based BDD Automation.}
Recent studies explore LLMs to automate BDD and acceptance testing.
\citet{Karpurapu2024} evaluated LLMs for generating Gherkin acceptance tests---referred as feature files in our work---and found that few-shot prompting can improve their quality.
\citet{Ferreira2025} used GPT-4 Turbo to generate Gherkin scenarios from user stories through AutoUAT, and Cypress tests from those scenarios and page HTML through Test Flow.
AutoUAT received positive feedback in 62 of 65 responses (95\%).
In a manual evaluation of 50 test cases from 13 user stories, 60\% of Test Flow's outputs accurately reflected the Gherkin scenarios, 8\% required minor fixes, 24\% required regeneration with additional context, and 8\% were discarded.
\citet{Paduraru2025} introduced BDDTestAIGen, which combines LLMs, NLP techniques, human feedback, and agentic mechanisms to create BDD tests by matching natural-language steps to existing implementations.
In a quantitative evaluation, the framework correctly matched 1,484 of 1,983 test variants generated by GPT-4 to the corresponding step implementations.

\noindentparagraph{Test Generation and Agentic Software Engineering.}
Most existing work on automated test generation supported by LLMs focuses on unit tests.
\citet{Schafer2024} introduced TestPilot, which uses execution feedback to iteratively generate and repair unit tests.
Using GPT-3.5 Turbo on 25 JavaScript npm packages comprising 1,684 API functions, TestPilot achieved median statement and branch coverage of 70.2\% and 52.8\%, compared with 51.3\% and 25.6\% for Nessie.
Similarly, \citet{Chen2024ChatUniTest} introduced ChatUniTest, which combines adaptive context selection with a generation-validation-repair process.
An evaluation on 264 methods from four Java projects reported overall line coverage of 59.6\%, compared with 42.1\% for TestSpark and 38.2\% for EvoSuite.
However, each baseline failed to generate tests for one project because of token limits or JDK incompatibility, respectively.

Agentic software engineering extends this automation to tasks that require interaction with development environments.
SWE-agent~\cite{yang2024sweagentagentcomputerinterfacesenable} introduces an agent with a purpose-built interface for navigating repositories, editing code, and executing programs to resolve software issues.
With GPT-4 Turbo, it resolved 12.47\% of SWE-bench test tasks and achieved a pass@1 rate of 87.7\% on the Python tasks of HumanEvalFix.
MetaGPT~\cite{hong2024metagptmetaprogrammingmultiagent} coordinates specialized agents through standard operating procedures and structured artifacts for requirements analysis, design, implementation, and testing.
It also uses execution feedback for debugging and, with GPT-4, achieved pass@1 rates of 85.9\% on HumanEval and 87.7\% on MBPP.

\noindentparagraph{\textbf{The role of BrinAI.}}
Our work is positioned at the intersection of BDD automation, LLM-based test generation, and agentic software engineering.
BrinAI is designed to generate step definitions from already specified feature files, thus enabling BDD scenarios to be automatically executed and validated against the system under test.
As BrinAI relies on an agentic architecture, it can plan, explore the project context, execute candidate implementations, and iteratively repair them based on observed failures.

\section{Conclusion}
\label{sec:conclusion}

In this paper, we proposed BrinAI, an agent-based approach that automates BDD scenario execution by generating step definition files.
Its workflow combines repository exploration, implementation planning, code generation, test execution, and iterative repair.
We also introduced GWT-Verified and GWT-Light, two fully executable datasets with 713 and 171 BDD snapshots, respectively. The snapshots were extracted from open-source projects hosted on GitHub and cover three programming languages: Java, Ruby, and Node.js.

Overall, BrinAI successfully executed 77 of the 171 GWT-Light snapshots (45.0\%), while the best prompt engineering strategy (zero-shot) executed 22 (12.9\%).
This difference remains regardless of the programming language, with BrinAI outperforming the LLM-baseline by 27.3, 31.6, and 33.7 p.p. in Java, Node.js, and Ruby, respectively.
In terms of median \textit{reproduction recall}, BrinAI outperformed the zero-shot baseline in Java (0.549 vs.\ 0.325) and Ruby (0.333 vs.\ 0.200), but achieved lower recall in Node.js (0.176 vs.\ 0.333).
As expected, BrinAI consumed more tokens during the execution process, an average of 460,475 tokens per snapshot, while the LLM-baseline consumed 4,253.
This difference directly impacted the cost of our approach, with BrinAI costing 18.5 times more than the LLM-baseline, at (US\$0.0973 vs. US\$0.0053, per snapshot).

\noindentparagraph{Future Work.}
Our next steps include (a) extending BrinAI's support to additional BDD frameworks, including Behave, (b) supporting additional programming languages, and (c) evaluating BrinAI on a larger set of BDD scenarios, including those from industrial projects.
We also plan to
(d) conduct a controlled experiment with developers to assess the quality of the generated step definitions, and in which situations they are more effective than developer-written ones.
We also intend to (e) investigate other interaction strategies, such as human-in-the-loop approaches, to improve the quality of the generated step definitions.
Finally, we intend to (f) compare BrinAI's multi-agent workflow with a single agent using the same model, tools, execution feedback, and token budget.

\noindentparagraph{Replication Package.}
BrinAI's source code is openly available at \url{https://github.com/brin-ai/brin-ai}.
Both GWT-Verified and GWT-Light datasets are available at \url{https://doi.org/10.5281/zenodo.24096530}.
Finally, the replication package for our experiments is available at \url{https://doi.org/10.5281/zenodo.24097200}.

\begin{acks}
  This work was partially supported by INES.IA (National Institute of Science and Technology for Software Engineering Based on and for Artificial Intelligence) \url{www.ines.org.br}, CNPq grant 408817/2024-0.
  It was also supported by grants from CNPq (403304/2025-3, 444507/2024-8) and FAPEMIG (APQ-02419-23).
\end{acks}

\bibliographystyle{ACM-Reference-Format}
\bibliography{bibliography}

@book{Wynne2017,
  title     = {The {{Cucumber Book}}: {{Behaviour-Driven Development}} for {{Testers}} and {{Developers}}},
  author    = {Wynne, Matt and Hellesoy, Aslak and Tooke, Steve},
  year      = {2017},
  publisher = {Pragmatic Bookshelf},
  address   = {Raleigh, NC, USA}
}

@book{Evans2003,
  title     = {Domain-{{Driven Design}}: {{Tackling Complexity}} in the {{Heart}} of {{Software}}},
  author    = {Evans, Eric},
  year      = {2003},
  publisher = {Addison-Wesley Professional},
  address   = {Boston, MA, USA}
}

@article{Schafer2024,
  title   = {An {{Empirical Evaluation}} of {{Using Large Language Models}} for {{Automated Unit Test Generation}}},
  author  = {Sch{\"a}fer, Max and Nadi, Sarah and Eghbali, Aryaz and Tip, Frank},
  year    = {2024},
  month   = jan,
  journal = {IEEE Transactions on Software Engineering},
  volume  = {50},
  number  = {1},
  pages   = {85--105}
}

@book{Winters2020,
  title     = {{Software Engineering at Google: Lessons Learned from Programming Over Time}},
  author    = {Winters, Titus and Manshreck, Tom and Wright, Hyrum},
  year      = {2020},
  edition   = {1st Edition},
  publisher = {O'Reilly Media},
  address   = {Sebastopol, CA, USA}
}

@inproceedings{Meglio2025,
  title     = {{{E2EGit}}: {{A Dataset}} of {{End-to-End Web Tests}} in {{Open Source Projects}}},
  booktitle = {22nd {{International Conference}} on {{Mining Software Repositories}} ({{MSR}})},
  author    = {Meglio, Sergio Di and Starace, Luigi Libero Lucio and Pontillo, Valeria and Opdebeeck, Ruben and Roover, Coen De and Martino, Sergio Di},
  year      = 2025,
  address   = {Ottawa, Ontario, Canada},
  publisher = {IEEE},
  pages     = {836--840}
}

@inproceedings{Just2014,
  title      = {{{Defects4J}}: A Database of Existing Faults to Enable Controlled Testing Studies for {{Java}} Programs},
  shorttitle = {{{Defects4J}}},
  booktitle  = {2014 {{International Symposium}} on {{Software Testing}} and {{Analysis}} ({{ISSTA}})},
  author     = {Just, Ren{\'e} and Jalali, Darioush and Ernst, Michael D.},
  year       = 2014,
  publisher = {Association for Computing Machinery},
  address   = {New York, NY, USA},
  pages      = {437--440}
}

@inproceedings{GivenWhenThenDataset,
  title     = {GivenWhenThen: A Dataset of BDD Test Scenarios Mined from Open Source Projects},
  author    = {Luciano Belo and João Montandon},
  booktitle = {23nd {{International Conference}} on {{Mining Software Repositories}} (MSR 2026)},
  year      = {2026},
  pages     = {595--599},
  publisher = {ACM},
  address   = {Rio de Janeiro, Brazil},
  doi       = {10.1145/3793302.3793308}
}

@article{binamungu2023bdd,
  author    = {Binamungu, Leonard Peter and Maro, Shem},
  title     = {Behaviour driven development: A systematic mapping study},
  journal   = {Journal of Systems and Software},
  volume    = {203},
  pages     = {111749},
  year      = {2023},
  doi       = {10.1016/j.jss.2023.111749}
}

@article{munaiah2017curating,
  author    = {Munaiah, Nuthan and Kroh, Steven and Cabrey, Craig and Nagappan, Meiyappan},
  title     = {Curating {GitHub} for engineered software projects},
  journal   = {Empirical Software Engineering},
  volume    = {22},
  number    = {6},
  pages     = {3219--3253},
  year      = {2017},
  doi       = {10.1007/s10664-017-9512-6}
}

@inproceedings{kalliamvakou2014promises,
  author    = {Kalliamvakou, Eirini and Gousios, Georgios and Blincoe, Kelly and Singer, Leif and German, Daniel M. and Damian, Daniela},
  title     = {The Promises and Perils of Mining {GitHub}},
  booktitle = {Proceedings of the 11th Working Conference on Mining Software Repositories},
  year      = {2014},
  pages     = {92--101},
  publisher = {ACM},
  address   = {Hyderabad, India},
  doi       = {10.1145/2597073.2597074}
}

@article{boettiger2015docker,
  author    = {Boettiger, Carl},
  title     = {An introduction to {Docker} for reproducible research, with examples from the {R} environment},
  journal   = {ACM SIGOPS Operating Systems Review},
  volume    = {49},
  number    = {1},
  pages     = {71--79},
  year      = {2015},
  doi       = {10.1145/2723872.2723882}
}

@inproceedings{lewis2021retrievalaugmentedgenerationknowledgeintensivenlp,
    author = {Lewis, Patrick and Perez, Ethan and Piktus, Aleksandra and Petroni, Fabio and Karpukhin, Vladimir and Goyal, Naman and K{\"u}ttler, Heinrich and Lewis, Mike and Yih, Wen-tau and Rockt{\"a}schel, Tim and Riedel, Sebastian and Kiela, Douwe},
    title = {Retrieval-augmented generation for knowledge-intensive NLP tasks},
    year = {2020},
    isbn = {9781713829546},
    publisher = {Curran Associates Inc.},
    address = {Red Hook, NY, USA},
    booktitle = {Proceedings of the 34th International Conference on Neural Information Processing Systems},
    articleno = {793},
    numpages = {16},
    location = {Vancouver, BC, Canada},
    series = {NIPS '20}
}

@inproceedings{reimers2019sentencebertsentenceembeddingsusing,
  author       = {Nils Reimers and Iryna Gurevych},
  editor       = {Kentaro Inui and Jing Jiang and Vincent Ng and Xiaojun Wan},
  title        = {Sentence-BERT: Sentence Embeddings using Siamese BERT-Networks},
  booktitle    = {Proceedings of the 2019 Conference on Empirical Methods in Natural
                  Language Processing and the 9th International Joint Conference on
                  Natural Language Processing, {EMNLP-IJCNLP} 2019, Hong Kong, China,
                  November 3-7, 2019},
  pages        = {3980--3990},
  publisher    = {Association for Computational Linguistics},
  address = {Hong Kong, China},
  year         = {2019},
  url          = {https://doi.org/10.18653/v1/D19-1410},
  doi          = {10.18653/V1/D19-1410},
  bibsource    = {dblp computer science bibliography, https://dblp.org}
}

@inproceedings{feng2020codebertpretrainedmodelprogramming,
    title = "{C}ode{BERT}: A Pre-Trained Model for Programming and Natural Languages",
    author = "Feng, Zhangyin  and
      Guo, Daya  and
      Tang, Duyu  and
      Duan, Nan  and
      Feng, Xiaocheng  and
      Gong, Ming  and
      Shou, Linjun  and
      Qin, Bing  and
      Liu, Ting  and
      Jiang, Daxin  and
      Zhou, Ming",
    editor = "Cohn, Trevor  and He, Yulan  and Liu, Yang",
    booktitle = "Findings of the {A}ssociation for {C}omputational {L}inguistics: {EMNLP} 2020",
    month = nov,
    year = "2020",
    address = "Online",
    publisher = "Association for Computational Linguistics",
    url = "https://aclanthology.org/2020.findings-emnlp.139/",
    doi = "10.18653/v1/2020.findings-emnlp.139",
    pages = "1536--1547"
}

@misc{Chen2021EvaluatingLL,
  title={Evaluating Large Language Models Trained on Code},
  author={Mark Chen and Jerry Tworek and Heewoo Jun and Qiming Yuan and Henrique Pond{\'e} and Jared Kaplan and Harrison Edwards and Yura Burda and Nicholas Joseph and Greg Brockman and Alex Ray and Raul Puri and Gretchen Krueger and Michael Petrov and Heidy Khlaaf and Girish Sastry and Pamela Mishkin and Brooke Chan and Scott Gray and Nick Ryder and Mikhail Pavlov and Alethea Power and Lukasz Kaiser and Mo Bavarian and Clemens Winter and Phil Tillet and Felipe Petroski Such and David W. Cummings and Matthias Plappert and Fotios Chantzis and Elizabeth Barnes and Ariel Herbert-Voss and William H. Guss and Alex Nichol and Igor Babuschkin and Suchir Balaji and Shantanu Jain and Andrew Carr and Jan Leike and Josh Achiam and Vedant Misra and Evan Morikawa and Alec Radford and Matthew M. Knight and Miles Brundage and Mira Murati and Katie Mayer and Peter Welinder and Bob McGrew and Dario Amodei and Sam McCandlish and Ilya Sutskever and Wojciech Zaremba},
  journal={ArXiv},
  year={2021},
  volume={abs/2107.03374},
  pages={1--35},
  numpages={35},
  url={https://api.semanticscholar.org/CorpusID:235755472}
}

@inproceedings{cambronero2019deeplearningmetcode,
    author = {Cambronero, Jose and Li, Hongyu and Kim, Seohyun and Sen, Koushik and Chandra, Satish},
    title = {When deep learning met code search},
    year = {2019},
    isbn = {9781450355728},
    publisher = {Association for Computing Machinery},
    address = {New York, NY, USA},
    url = {https://doi.org/10.1145/3338906.3340458},
    doi = {10.1145/3338906.3340458},
    booktitle = {Proceedings of the 2019 27th ACM Joint Meeting on European Software Engineering Conference and Symposium on the Foundations of Software Engineering},
    pages = {964–974},
    numpages = {11},
    location = {Tallinn, Estonia},
    series = {ESEC/FSE 2019}
}

@inproceedings{wei2023chainofthoughtpromptingelicitsreasoning,
    author = {Wei, Jason and Wang, Xuezhi and Schuurmans, Dale and Bosma, Maarten and Ichter, Brian and Xia, Fei and Chi, Ed H. and Le, Quoc V. and Zhou, Denny},
    title = {Chain-of-thought prompting elicits reasoning in large language models},
    year = {2022},
    isbn = {9781713871088},
    publisher = {Curran Associates Inc.},
    address = {Red Hook, NY, USA},
    booktitle = {Proceedings of the 36th International Conference on Neural Information Processing Systems},
    articleno = {1800},
    numpages = {14},
    location = {New Orleans, LA, USA},
    series = {NIPS '22}
}

@article{liu2021pretrainpromptpredictsystematic,
    author = {Liu, Pengfei and Yuan, Weizhe and Fu, Jinlan and Jiang, Zhengbao and Hayashi, Hiroaki and Neubig, Graham},
    title = {Pre-train, Prompt, and Predict: A Systematic Survey of Prompting Methods in Natural Language Processing},
    year = {2023},
    issue_date = {September 2023},
    publisher = {Association for Computing Machinery},
    address = {New York, NY, USA},
    volume = {55},
    number = {9},
    issn = {0360-0300},
    url = {https://doi.org/10.1145/3560815},
    doi = {10.1145/3560815},
    journal = {ACM Comput. Surv.},
    month = jan,
    articleno = {195},
    numpages = {35}
}

@article{Wang_2024,
   title={A survey on large language model based autonomous agents},
   volume={18},
   ISSN={2095-2236},
   url={http://dx.doi.org/10.1007/s11704-024-40231-1},
   DOI={10.1007/s11704-024-40231-1},
   number={6},
   pages={186345},
   journal={Frontiers of Computer Science},
   publisher={Springer Science and Business Media LLC},
   author={Wang, Lei and Ma, Chen and Feng, Xueyang and Zhang, Zeyu and Yang, Hao and Zhang, Jingsen and Chen, Zhiyuan and Tang, Jiakai and Chen, Xu and Lin, Yankai and Zhao, Wayne Xin and Wei, Zhewei and Wen, Jirong},
   year={2024},
   month=Mar
}

@inproceedings{yang2024sweagentagentcomputerinterfacesenable,
    author = {Yang, John and Jimenez, Carlos E. and Wettig, Alexander and Lieret, Kilian and Yao, Shunyu and Narasimhan, Karthik and Press, Ofir},
    title = {SWE-agent: agent-computer interfaces enable automated software engineering},
    year = {2024},
    isbn = {9798331314385},
    publisher = {Curran Associates Inc.},
    address = {Red Hook, NY, USA},
    booktitle = {Proceedings of the 38th International Conference on Neural Information Processing Systems},
    articleno = {1601},
    numpages = {125},
    location = {Vancouver, BC, Canada},
    series = {NIPS '24}
}

@inproceedings{hong2024metagptmetaprogrammingmultiagent,
 author = {Hong, Sirui and Zhuge, Mingchen and Chen, Jonathan and Zheng, Xiawu and Cheng, Yuheng and Wang, Jinlin and Zhang, Ceyao and wang, zili and Yau, Steven and Lin, Zijuan and Zhou, Liyang and Ran, Chenyu and Xiao, Lingfeng and Wu, Chenglin and Schmidhuber, J\"{u}rgen},
 booktitle = {International Conference on Learning Representations},
 editor = {B. Kim and Y. Yue and S. Chaudhuri and K. Fragkiadaki and M. Khan and Y. Sun},
 pages = {23247--23275},
 title = {MetaGPT: Meta Programming for A Multi-Agent Collaborative Framework},
 url = {https://proceedings.iclr.cc/paper_files/paper/2024/file/6507b115562bb0a305f1958ccc87355a-Paper-Conference.pdf},
 volume = {2024},
 year = {2024},
 publisher = {International Conference on Learning Representations},
 address   = {Vienna, Austria}
}

@InProceedings{10.1007/978-3-642-30561-0_19,
    author="Soeken, Mathias
    and Wille, Robert
    and Drechsler, Rolf",
    editor="Furia, Carlo A.
    and Nanz, Sebastian",
    title="Assisted Behavior Driven Development Using Natural Language Processing",
    booktitle="Objects, Models, Components, Patterns",
    year="2012",
    publisher="Springer Berlin Heidelberg",
    address="Berlin, Heidelberg",
    pages="269--287",
    isbn="978-3-642-30561-0"
}

@article{yang2025requirementsbasedtestgenerationcomprehensive,
    author = {Yang, Zhenzhen and Huang, Rubing and Cui, Chenhui and Niu, Nan and Towey, Dave},
    title = {Requirements-Based Test Generation: A Comprehensive Survey},
    year = {2026},
    issue_date = {October 2026},
    publisher = {Association for Computing Machinery},
    address = {New York, NY, USA},
    volume = {35},
    number = {10},
    issn = {1049-331X},
    url = {https://doi.org/10.1145/3771727},
    doi = {10.1145/3771727},
    month = sep,
    articleno = {343},
    numpages = {61},
    journal = {ACM Transactions on Software Engineering and Methodology}
}

@inproceedings{381473.381476,
author = {Ricca, Filippo and Tonella, Paolo},
title = {Analysis and testing of Web applications},
year = {2001},
isbn = {0769510507},
publisher = {IEEE Computer Society},
address = {USA},
booktitle = {Proceedings of the 23rd International Conference on Software Engineering},
pages = {25–34},
numpages = {10},
location = {Toronto, Ontario, Canada},
series = {ICSE '01}
}

@article{zhang2025patchempoweringlargelanguage,
    author = {Zhang, Yuwei and Jin, Zhi and Xing, Ying and Li, Ge and Liu, Fang and Zhu, Jiaxin and Dou, Wensheng and Wei, Jun},
    title = {PATCH: Empowering Large Language Model with Programmer-Intent Guidance and Collaborative-Behavior Simulation for Automatic Bug Fixing},
    year = {2025},
    issue_date = {January 2026},
    publisher = {Association for Computing Machinery},
    address = {New York, NY, USA},
    volume = {35},
    number = {1},
    issn = {1049-331X},
    url = {https://doi.org/10.1145/3718739},
    doi = {10.1145/3718739},
    month = dec,
    articleno = {3},
    numpages = {35},
    journal = {ACM Transactions on Software Engineering and Methodology}
}

@inproceedings{yao2023reactsynergizingreasoningacting,
  author    = {Yao, Shunyu and Zhao, Jeffrey and Yu, Dian and Du, Nan and Shafran, Izhak and Narasimhan, Karthik and Cao, Yuan},
  title     = {{ReAct}: Synergizing Reasoning and Acting in Language Models},
  booktitle = {The Eleventh International Conference on Learning Representations},
  year      = {2023},
  publisher = {OpenReview.net},
  address   = {[S. l.]},
  numpages = {TOTAL},
  url       = {https://openreview.net/forum?id=WE_vluYUL-X}
}

@book{huyen_ai_2025,
	address = {Sebastopol,CA},
	edition = {First edition},
	title = {{AI} engineering: building applications with foundation models},
	isbn = {978-1-0981-6630-4 978-1-0981-6627-4},
	shorttitle = {{AI} engineering},
	language = {en},
	publisher = {O'Reilly},
	author = {Huyen, Chip},
	year = {2025},
}

@inproceedings{luo2014empirical,
  author    = {Luo, Qingzhou and Hariri, Farah and Eloussi, Lamyaa and Marinov, Darko},
  title     = {An Empirical Analysis of Flaky Tests},
  booktitle = {Proceedings of the 22nd ACM SIGSOFT International Symposium on Foundations of Software Engineering (FSE)},
  year      = {2014},
  pages     = {643--653},
  publisher = {ACM},
  address   = {Hong Kong, China},
  doi       = {10.1145/2635868.2635920}
}

@inproceedings{Nagappan2013Diversity,
    author = {Nagappan, Meiyappan and Zimmermann, Thomas and Bird, Christian},
    title = {Diversity in software engineering research},
    year = {2013},
    isbn = {9781450322379},
    publisher = {Association for Computing Machinery},
    address = {New York, NY, USA},
    url = {https://doi.org/10.1145/2491411.2491415},
    doi = {10.1145/2491411.2491415},
    booktitle = {Proceedings of the 2013 9th Joint Meeting on Foundations of Software Engineering},
    pages = {466–476},
    numpages = {11},
    location = {Saint Petersburg, Russia},
    series = {ESEC/FSE 2013}
}

@inproceedings{jimenez2024swebench,
 author = {Jimenez, Carlos E and Yang, John and Wettig, Alexander and Yao, Shunyu and Pei, Kexin and Press, Ofir and Narasimhan, Karthik},
 booktitle = {International Conference on Learning Representations},
 editor = {B. Kim and Y. Yue and S. Chaudhuri and K. Fragkiadaki and M. Khan and Y. Sun},
 publisher = {International Conference on Learning Representations (ICLR)},
 address   = {Appleton, WI, USA},
 pages = {54107--54157},
 title = {SWE-bench: Can Language Models Resolve Real-world Github Issues?},
 url = {https://proceedings.iclr.cc/paper_files/paper/2024/file/edac78c3e300629acfe6cbe9ca88fb84-Paper-Conference.pdf},
 volume = {2024},
 year = {2024}
}

@misc{amatriain2024promptdesignengineeringintroduction,
      title={Prompt Design and Engineering: Introduction and Advanced Methods}, 
      author={Xavier Amatriain},
      year={2024},
      eprint={2401.14423},
      archivePrefix={arXiv},
      primaryClass={cs.SE},
      url={https://arxiv.org/abs/2401.14423}, 
}

@article{Zampetti2020,
  title   = {Demystifying the adoption of behavior-driven development in open source projects},
  author  = {Zampetti, Fiorella and Di Sorbo, Andrea and Visaggio, Corrado Aaron and Canfora, Gerardo and Di Penta, Massimiliano},
  journal = {Information and Software Technology},
  volume  = {123},
  pages   = {106311},
  year    = {2020},
  doi     = {10.1016/j.infsof.2020.106311}
}

@inproceedings{Binamungu2018,
  title     = {Maintaining behaviour driven development specifications: Challenges and opportunities},
  author    = {Binamungu, Leonard Peter and Embury, Suzanne Marie and Konstantinou, Nikolaos},
  booktitle = {2018 IEEE 25th International Conference on Software Analysis, Evolution and Reengineering (SANER)},
  pages     = {175--184},
  year      = {2018},
  publisher = {IEEE},
  address   = {Campobasso, Italy},
  doi       = {10.1109/SANER.2018.8330207}
}

@article{Karpurapu2024,
    title={Comprehensive Evaluation and Insights Into the Use of Large Language Models in the Automation of Behavior-Driven Development Acceptance Test Formulation},
   volume={12},
   ISSN={2169-3536},
   url={http://dx.doi.org/10.1109/ACCESS.2024.3391815},
   DOI={10.1109/access.2024.3391815},
   journal={IEEE Access},
   publisher={Institute of Electrical and Electronics Engineers (IEEE)},
   author={Karpurapu, Shanthi and Myneni, Sravanthy and Nettur, Unnati and Gajja, Likhit Sagar and Burke, Dave and Stiehm, Tom and Payne, Jeffery},
   year={2024},
   pages={58715–58721}
}

@inproceedings{Ferreira2025,
  author    = {Ferreira, Margarida and Viegas, Luís and Faria, João Pascoal and Lima, Bruno},
  title     = {Acceptance Test Generation with Large Language Models: An Industrial Case Study},
  booktitle = {2025 IEEE/ACM International Conference on Automation of Software Test (AST)},
  year      = {2025},
  month     = apr,
  pages     = {1--11},
  address   = {Ottawa, ON, Canada},
  publisher = {IEEE},
  doi       = {10.1109/AST66626.2025.00007},
  url       = {https://doi.org/10.1109/AST66626.2025.00007}
}

@inproceedings{Paduraru2025,
  booktitle = {17th International Conference on Agents and Artificial IntelligenceAt: ICAART'25Volume: 2},
  author = {Paduraru, Ciprian and Zavelca, Miruna and Stefanescu, Alin},
  year = {2025},
  month = {02},
  pages = {805-815},
  title = {Agentic AI for Behavior-Driven Development Testing Using Large Language Models},
  doi = {10.5220/0013374400003890},
  publisher = {SciTePress},
  address = {Porto, Portugal}
}

@inproceedings{Chen2024ChatUniTest,
    author = {Chen, Yinghao and Hu, Zehao and Zhi, Chen and Han, Junxiao and Deng, Shuiguang and Yin, Jianwei},
    title = {ChatUniTest: A Framework for LLM-Based Test Generation},
    year = {2024},
    isbn = {9798400706585},
    publisher = {Association for Computing Machinery},
    address = {New York, NY, USA},
    url = {https://doi.org/10.1145/3663529.3663801},
    doi = {10.1145/3663529.3663801},
    booktitle = {Companion Proceedings of the 32nd ACM International Conference on the Foundations of Software Engineering},
    pages = {572–576},
    numpages = {5},
    location = {Porto de Galinhas, Brazil},
    series = {FSE 2024}
}

@incollection{basili1994gqm,
  author    = {Basili, Victor R. and Caldiera, Gianluigi and Rombach, H. Dieter},
  title     = {The Goal Question Metric Approach},
  booktitle = {Encyclopedia of Software Engineering},
  editor    = {Marciniak, John J.},
  publisher = {John Wiley \& Sons},
  address   = {New York, NY, USA},
  pages     = {528--532},
  year      = {1994}
}

@article{arcuri2014hitchhiker,
  author    = {Arcuri, Andrea and Briand, Lionel},
  title     = {A Hitchhiker's Guide to Statistical Tests for Assessing Randomized Algorithms in Software Engineering},
  journal   = {Software Testing, Verification and Reliability},
  volume    = {24},
  number    = {3},
  pages     = {219--250},
  month     = may,
  year      = {2014},
  publisher = {John Wiley \& Sons, Ltd.},
  issn      = {0960-0833},
  doi       = {10.1002/stvr.1486},
  url       = {https://doi.org/10.1002/stvr.1486}
}

@inproceedings{bouzenia2025repairagent,
  author    = {Bouzenia, Islem and Devanbu, Premkumar and Pradel, Michael},
  title     = {{RepairAgent}: An Autonomous, {LLM}-Based Agent for Program Repair},
  booktitle = {2025 IEEE/ACM 47th International Conference on Software Engineering (ICSE)},
  year      = {2025},
  pages     = {2188--2200},
  publisher = {IEEE},
  doi       = {10.1109/ICSE55347.2025.00157},
  address   = {Ottawa, ON, Canada},
  url       = {https://doi.org/10.1109/ICSE55347.2025.00157}
}

@inproceedings{gou2024critic,
  author    = {Gou, Zhibin and Shao, Zhihong and Gong, Yeyun and Shen, Yelong and Yang, Yujiu and Duan, Nan and Chen, Weizhu},
  title     = {{CRITIC}: Large Language Models Can Self-Correct with Tool-Interactive Critiquing},
  booktitle = {The Twelfth International Conference on Learning Representations},
  year      = {2024},
  publisher = {OpenReview.net},
  address   = {Vienna, Austria},
  numpages  = {78},
  url       = {https://openreview.net/forum?id=Sx038qxjek}
}

@inproceedings{ma2025thinking,
    author = {Ma, Yingwei and Li, Yongbin and Dong, Yihong and Jiang, Xue and Li, Yanhao and Liu, Yue and Cao, Rongyu and Chen, Jue and Huang, Fei and Li, Binhua},
    title = {Thinking Longer, Not Larger: Enhancing Software Engineering Agents via Scaling Test-Time Compute},
    year = {2025},
    publisher = {IEEE Press},
    url = {https://doi.org/10.1109/ASE63991.2025.00309},
    doi = {10.1109/ASE63991.2025.00309},
    booktitle = {2025 40th IEEE/ACM International Conference on Automated Software Engineering (ASE)},
    pages = {3730–-3741},
    address = {Piscataway, NJ, USA},
    numpages = {12}
}

@inproceedings{nijkamp2023codegen,
  author    = {Nijkamp, Erik and Pang, Bo and Hayashi, Hiroaki and Tu, Lifu and Wang, Huan and Zhou, Yingbo and Savarese, Silvio and Xiong, Caiming},
  title     = {{CodeGen}: An Open Large Language Model for Code with Multi-Turn Program Synthesis},
  booktitle = {International Conference on Learning Representations},
  year      = {2023},
  publisher = {OpenReview.net},
  address   = {Kigali, Rwanda},
  numpages = {25},
  url       = {https://openreview.net/forum?id=iaYcJKpY2B_}
}

@inproceedings{xia2023repair,
  author    = {Xia, Chunqiu Steven and Wei, Yuxiang and Zhang, Lingming},
  title     = {Automated Program Repair in the Era of Large
               Pre-trained Language Models},
  booktitle = {2023 IEEE/ACM 45th International Conference on Software Engineering (ICSE)},
  address   = {Melbourne, Australia},
  publisher = {IEEE},
  year      = {2023},
  pages     = {1482--1494},
  doi       = {10.1109/ICSE48619.2023.00129}
}

@article{Wang2022,
  author  = {Wang, Yuqing and Mäntylä, Mika V.
             and Liu, Zihao and Markkula, Jouni},
  title   = {Test Automation Maturity Improves Product
             Quality: Quantitative Study of Open Source
             Projects Using Continuous Integration},
  journal = {Journal of Systems and Software},
  volume  = {188},
  pages   = {111259},
  year    = {2022},
  doi     = {10.1016/j.jss.2022.111259}
}

@article{Shahin2017,
  author  = {Shahin, Mojtaba and Babar, Muhammad Ali and Zhu, Liming},
  title   = {Continuous Integration, Delivery and Deployment: A Systematic Review on Approaches, Tools, Challenges and Practices},
  journal = {IEEE Access},
  volume  = {5},
  pages   = {3909--3943},
  year    = {2017},
  doi     = {10.1109/ACCESS.2017.2685629}
}

@book{Humble2010,
  author    = {Humble, Jez and Farley, David},
  title     = {Continuous Delivery: Reliable Software Releases through Build, Test, and Deployment Automation},
  publisher = {Addison-Wesley Professional},
  year      = {2010},
  address   = {Upper Saddle River, NJ, USA},
  isbn      = {9780321601919}
}

@misc{tdd_fowler,
  author       = {Fowler, Martin},
  title        = {Test Driven Development},
  year         = {2023},
  howpublished = {\url{https://martinfowler.com/bliki/TestDrivenDevelopment.html}},
  note         = {Originally published March 5, 2005;
                  updated December 11, 2023}
}

@inproceedings{Godefroid2005,
  author = {Godefroid, Patrice and Klarlund, Nils and Sen, Koushik},
  title = {DART: Directed Automated Random Testing},
  booktitle = {Proceedings of the ACM SIGPLAN 2005 Conference on Programming Language Design and Implementation},
  pages = {213--223},
  year = {2005},
  publisher = {ACM},
  address = {New York, NY, USA},
  doi = {10.1145/1065010.1065036}
}

@inproceedings{Pacheco2007,
  author = {Pacheco, Carlos and Lahiri, Shuvendu K. and Ernst, Michael D. and Ball, Thomas},
  title = {Feedback-Directed Random Test Generation},
  booktitle = {29th International Conference on Software Engineering (ICSE)},
  pages = {75--84},
  year = {2007},
  publisher = {IEEE},
  address   = {Minneapolis, MN, USA},
  doi = {10.1109/ICSE.2007.37}
}

@inproceedings{Fraser2011,
  author = {Fraser, Gordon and Arcuri, Andrea},
  title = {EvoSuite: Automatic Test Suite Generation for Object-Oriented Software},
  booktitle = {Proceedings of the 19th ACM SIGSOFT Symposium and the 13th European Conference on Foundations of Software Engineering},
  pages = {416--419},
  year = {2011},
  publisher = {ACM},
  address   = {New York, NY, USA},
  doi = {10.1145/2025113.2025179}
}

@inproceedings{Wiecher2021,
  author = {Carsten Wiecher and Jannik Fischbach and Joel Greenyer and Andreas Vogelsang and Carsten Wolff and Roman Dumitrescu},
  title = {Integrated and Iterative Requirements Analysis and Test Specification: A Case Study at Kostal},
  booktitle = {2021 ACM/IEEE 24th International Conference on Model Driven Engineering Languages and Systems (MODELS)},
  year = {2021},
  pages = {112--122},
  address   = {Fukuoka, Japan},
  publisher = {IEEE},
  doi = {10.1109/MODELS50736.2021.00020}
}

@inproceedings{Liu2024RepoBench,
 author = {Liu, Tianyang and Xu, Canwen and McAuley, Julian },
 booktitle = {International Conference on Learning Representations},
 editor = {B. Kim and Y. Yue and S. Chaudhuri and K. Fragkiadaki and M. Khan and Y. Sun},
 publisher = {International Conference on Learning Representations (ICLR)},
 address   = {Appleton, WI, USA},
 pages = {47832--47850},
 title = {RepoBench: Benchmarking Repository-Level Code Auto-Completion Systems},
 url = {https://proceedings.iclr.cc/paper_files/paper/2024/file/d191ba4c8923ed8fd8935b7c98658b5f-Paper-Conference.pdf},
 volume = {2024},
 year = {2024}
}

@inproceedings{Zhang2023RepoCoder,
  author = {Fengji Zhang and Bei Chen and Yue Zhang and Jacky Keung and Jin Liu and Daoguang Zan and Yi Mao and Jian-Guang Lou and Weizhu Chen},
  title = {RepoCoder: Repository-Level Code Completion Through Iterative Retrieval and Generation},
  booktitle = {Proceedings of EMNLP},
  pages = {2471--2484},
  year = {2023},
  address = {Singapore},
  publisher = {Association for Computational Linguistics},
  doi = {10.18653/v1/2023.emnlp-main.151}
}

@inproceedings{Wang2025CodeRAGBench,
  author = {Zora Zhiruo Wang and Akari Asai and Xinyan Velocity Yu and Frank F. Xu and Yiqing Xie and Graham Neubig and Daniel Fried},
  title = {CodeRAG-Bench: Can Retrieval Augment Code Generation?},
  booktitle = {Findings of NAACL},
  address   = {Albuquerque, New Mexico},
  publisher = {Association for Computational Linguistics},
  pages = {3199--3214},
  year = {2025},
  doi = {10.18653/v1/2025.findings-naacl.176}
}

@inproceedings{Harman2024,
  author = {Nadia Alshahwan and Jubin Chheda and Anastasia Finogenova and Beliz Gokkaya and Mark Harman and Inna Harper and Alexandru Marginean and Shubho Sengupta and Eddy Wang},
  title = {Automated Unit Test Improvement using Large Language Models at Meta},
  booktitle = {Companion Proceedings of the 32nd ACM International Conference on the Foundations of Software Engineering},
  address   = {New York, NY, USA},
  publisher = {Association for Computing Machinery},
  pages = {185--196},
  year = {2024},
  doi = {10.1145/3663529.3663839}
}

@article{CoverUp2025,
  author = {Juan Altmayer Pizzorno and Emery D. Berger},
  title = {CoverUp: Effective High Coverage Test Generation for Python},
  journal = {Proceedings of the ACM on Software Engineering},
  volume = {2},
  number = {FSE},
  year = {2025},
  pages     = {2897--2919},
  articleno = {FSE128},
  numpages  = {23},
  doi = {10.1145/3729398}
}

@inproceedings{guelman2026quality,
  author    = {Guelman, Ian and Leal, Arthur Gregório and Xavier, Laerte and Valente, Marco Tulio},
  title     = {On the Quality of AI-Generated Source Code Comments: A Comprehensive Evaluation},
  booktitle = {Proceedings of the 1st International Workshop on AI for Software Quality Evaluation: Judgment, Metrics, Benchmarks, and Beyond (AI-SQE '26)},
  year      = {2026},
  month     = apr,
  publisher = {ACM},
  address   = {New York, NY, USA},
  numpages  = {9},
  doi       = {10.1145/3786175.3788344}
}

@inproceedings{trabalho_altino_ast,
author = {Alves, Altino and Montandon, Jo{\~a}o Eduardo and Hora, Andre},
title = {Testing Framework Migration with Large Language Models},
year = {2026},
isbn = {9798400724763},
publisher = {Association for Computing Machinery},
address = {New York, NY, USA},
url = {https://doi.org/10.1145/3793654.3793749},
doi = {10.1145/3793654.3793749},
booktitle = {Proceedings of the 7th ACM/IEEE International Conference on Automation of Software Test},
pages = {12–22},
numpages = {11},
location = {Rio de Janeiro, Brazil},
series = {AST '26}
}

@inproceedings{trabalho_aylton,
author = {Almeida, Aylton and Valente, Marco Tulio and Xavier, Laerte},
title = {Using Copilot Agent Mode to Automate Library Migration: A Quantitative Assessment},
year = {2026},
isbn = {9798400723995},
publisher = {Association for Computing Machinery},
address = {New York, NY, USA},
url = {https://doi.org/10.1145/3786167.3788411},
doi = {10.1145/3786167.3788411},
booktitle = {Proceedings of the 2026 International Workshop on Agentic Engineering},
pages = {135–139},
numpages = {5},
location = {Rio de Janeiro, Brazil},
series = {AGENT '26}
}

@inproceedings{trabalho_nadi_islam,
  author    = {Islam, Mohayeminul and Jha, Ajay Kumar and Mahmoud, May and Akhmetov, Ildar and Nadi, Sarah},
  title     = {An Empirical Study of Python Library Migration Using Large Language Models},
  booktitle = {2025 40th IEEE/ACM International Conference on Automated Software Engineering (ASE)},
  year      = {2025},
  pages     = {867--879},
  publisher = {IEEE},
  address   = {Seoul, South Korea},
  doi       = {10.1109/ASE63991.2025.00077}
}

@article{Islam_2024,
   title={Characterizing Python Library Migrations},
   volume={1},
   ISSN={2994-970X},
   url={http://dx.doi.org/10.1145/3643731},
   DOI={10.1145/3643731},
   number={FSE},
   journal={Proceedings of the ACM on Software Engineering},
   publisher={Association for Computing Machinery (ACM)},
   author={Islam, Mohayeminul and Jha, Ajay Kumar and Akhmetov, Ildar and Nadi, Sarah},
   year={2024},
   month={July}, 
   pages={92–114}
}

@inproceedings{du2024classeval,
  author    = {Du, Xueying and Liu, Mingwei and Wang, Kaixin and Wang, Hanlin and Liu, Junwei and Chen, Yixuan and Feng, Jiayi and Sha, Chaofeng and Peng, Xin and Lou, Yiling},
  title     = {Evaluating Large Language Models in Class-Level Code Generation},
  booktitle = {Proceedings of the 46th IEEE/ACM International Conference on Software Engineering (ICSE)},
  year      = {2024},
  pages     = {1--13},
  publisher = {ACM},
  address   = {New York, NY, USA},
  doi       = {10.1145/3597503.3639219}
}

@inproceedings{liu2023evalplus,
  author    = {Liu, Jiawei and Xia, Chunqiu Steven and Wang, Yuyao and Zhang, Lingming},
  title     = {Is Your Code Generated by {ChatGPT} Really Correct? Rigorous Evaluation of Large Language Models for Code Generation},
  booktitle = {Advances in Neural Information Processing Systems (NeurIPS)},
  volume    = {36},
  year      = {2023},
  pages     = {21558--21572},
  publisher = {Curran Associates, Inc.},
  address   = {Red Hook, NY, USA},
  url       = {https://proceedings.neurips.cc/paper_files/paper/2023/hash/43e9d647ccd3e4b7b5baab53f0368686-Abstract.html}
}

@inproceedings{wang2023codet5plus,
  author    = {Wang, Yue and Le, Hung and Gotmare, Akhilesh and Bui, Nghi and Li, Junnan and Hoi, Steven},
  title     = {{CodeT5+}: Open Code Large Language Models for Code Understanding and Generation},
  booktitle = {Proceedings of the 2023 Conference on Empirical Methods in Natural Language Processing (EMNLP)},
  year      = {2023},
  pages     = {1069--1088},
  publisher = {Association for Computational Linguistics},
  address   = {Singapore},
  doi       = {10.18653/v1/2023.emnlp-main.68}
}

@inproceedings{geng2024summarizers,
  author    = {Geng, Mingyang and Wang, Shangwen and Dong, Dezun and Wang, Haotian and Li, Ge and Jin, Zhi and Mao, Xiaoguang and Liao, Xiangke},
  title     = {Large Language Models are Few-Shot Summarizers: Multi-Intent Comment Generation via In-Context Learning},
  booktitle = {Proceedings of the 46th IEEE/ACM International Conference on Software Engineering (ICSE)},
  year      = {2024},
  pages     = {1--13},
  publisher = {ACM},
  address   = {New York, NY, USA},
  doi       = {10.1145/3597503.3608134}
}

@inproceedings{sun2025summarization,
  author    = {Sun, Weisong and Miao, Yun and Li, Yuekang and Zhang, Hongyu and Fang, Chunrong and Liu, Yi and Deng, Gelei and Liu, Yang and Chen, Zhenyu},
  title     = {Source Code Summarization in the Era of Large Language Models},
  booktitle = {Proceedings of the 47th IEEE/ACM International Conference on Software Engineering (ICSE)},
  year      = {2025},
  pages     = {1882--1894},
  publisher = {IEEE},
  address   = {Ottawa, ON, Canada},
  doi       = {10.1109/ICSE55347.2025.00034}
}

@article{bairi2024codeplan,
  author    = {Bairi, Ramakrishna and Sonwane, Atharv and Kanade, Aditya and {Vageesh D. C.} and Iyer, Arun and Parthasarathy, Suresh and Rajamani, Sriram and {B. Ashok} and Shet, Shashank},
  title     = {{CodePlan}: Repository-Level Coding using {LLMs} and Planning},
  journal   = {Proceedings of the ACM on Software Engineering},
  volume    = {1},
  number    = {FSE},
  year      = {2024},
  pages     = {675--698},
  doi       = {10.1145/3643757}
}

@inproceedings{lemieux2023codamosa,
  author    = {Lemieux, Caroline and Inala, Jeevana Priya and Lahiri, Shuvendu K. and Sen, Siddhartha},
  title     = {{CodaMOSA}: Escaping Coverage Plateaus in Test Generation with Pre-trained Large Language Models},
  booktitle = {Proceedings of the 45th IEEE/ACM International Conference on Software Engineering (ICSE)},
  year      = {2023},
  pages     = {919--931},
  publisher = {IEEE},
  address   = {Melbourne, Australia},
  doi       = {10.1109/ICSE48619.2023.00085}
}

@article{yuan2024chattester,
  author    = {Yuan, Zhiqiang and Liu, Mingwei and Ding, Shiji and Wang, Kaixin and Chen, Yixuan and Peng, Xin and Lou, Yiling},
  title     = {Evaluating and Improving {ChatGPT} for Unit Test Generation},
  journal   = {Proceedings of the ACM on Software Engineering},
  volume    = {1},
  number    = {FSE},
  year      = {2024},
  pages     = {1703--1726},
  doi       = {10.1145/3660783}
}

@misc{xia2024agentlessdemystifyingllmbasedsoftware,
      author = {Xia, Chunqiu Steven and Deng, Yinlin and Dunn, Soren and Zhang, Lingming},
    title = {Demystifying LLM-Based Software Engineering Agents},
    year = {2025},
    issue_date = {July 2025},
    publisher = {Association for Computing Machinery},
    address = {New York, NY, USA},
    volume = {2},
    number = {FSE},
    url = {https://doi.org/10.1145/3715754},
    doi = {10.1145/3715754},
    journal = {Proc. ACM Softw. Eng.},
    month = jun,
    articleno = {FSE037},
    numpages = {24}
}

@inproceedings{zhang2024autocoderoverautonomousprogramimprovement,
    author = {Zhang, Yuntong and Ruan, Haifeng and Fan, Zhiyu and Roychoudhury, Abhik},
    title = {AutoCodeRover: Autonomous Program Improvement},
    year = {2024},
    isbn = {9798400706127},
    publisher = {Association for Computing Machinery},
    address = {New York, NY, USA},
    url = {https://doi.org/10.1145/3650212.3680384},
    doi = {10.1145/3650212.3680384},
    booktitle = {Proceedings of the 33rd ACM SIGSOFT International Symposium on Software Testing and Analysis},
    pages = {1592–1604},
    numpages = {13},
    location = {Vienna, Austria},
    series = {ISSTA 2024}
}

@inproceedings{antoniades2025swesearch,
 author = {Antoniades, Antonis and \"{O}rwall, Albert and Zhang, Kexun and Xie, Yuxi and Goyal, Anirudh and Wang, William},
 booktitle = {International Conference on Learning Representations},
 editor = {Y. Yue and A. Garg and N. Peng and F. Sha and R. Yu},
 publisher = {International Conference on Learning Representations (ICLR)},
 address   = {Appleton, WI, USA},
 pages = {64485--64515},
 title = {SWE-Search: Enhancing Software Agents with Monte Carlo Tree Search and Iterative Refinement},
 url = {https://proceedings.iclr.cc/paper_files/paper/2025/file/a1e6783e4d739196cad3336f12d402bf-Paper-Conference.pdf},
 volume = {2025},
 year = {2025}
}

@inproceedings{kastner,
    author = {Nahar, Nadia and Zhou, Shurui and Lewis, Grace and K{\"a}stner, Christian},
    title = {Collaboration challenges in building ML-enabled systems: communication, documentation, engineering, and process},
    year = {2022},
    isbn = {9781450392211},
    publisher = {Association for Computing Machinery},
    address = {New York, NY, USA},
    url = {https://doi.org/10.1145/3510003.3510209},
    doi = {10.1145/3510003.3510209},
    booktitle = {Proceedings of the 44th International Conference on Software Engineering},
    pages = {413–425},
    numpages = {13},
    location = {Pittsburgh, Pennsylvania},
    series = {ICSE '22}
}

@inproceedings{jain2024mitigatingcodellmhallucinations,
    author = {Jain, Nihal and Kwiatkowski, Robert and Ray, Baishakhi and Ramanathan, Murali Krishna and Kumar, Varun},
    title = {On Mitigating Code LLM Hallucinations with API Documentation},
    year = {2025},
    publisher = {IEEE Press},
    address   = {Piscataway, NJ, USA},
    url = {https://doi.org/10.1109/ICSE-SEIP66354.2025.00027},
    doi = {10.1109/ICSE-SEIP66354.2025.00027},
    booktitle = {2025 IEEE/ACM 47th International Conference on Software Engineering: Software Engineering in Practice (ICSE-SEIP)},
    pages = {237–248},
    numpages = {12}
}

@inproceedings{helio_agents,
  author    = {dos Santos, Hélio Victor Flexa and Costa, Vitor and Montandon, João Eduardo and Valente, Marco Tulio},
  title     = {Decoding the Configuration of AI Coding Agents: Insights from Claude Code Projects},
  booktitle = {Proceedings of the 2026 International Workshop on Agentic Engineering (AGENT)},
  year      = {2026},
  pages     = {63--67},
  publisher = {ACM},
  address   = {New York, NY, USA},
  doi       = {10.1145/3786167.3788412}
}

@inproceedings{helio_scam,
  author    = {dos Santos, Hélio Victor Flexa and Costa, Vitor and Montandon, João Eduardo and Silva, Luciana Lourdes and Valente, Marco Tulio},
  title     = {Configuration Smells in {AGENTS.md} Files: Common Mistakes in Configuring Coding Agents},
  booktitle = {Proceedings of the 26th IEEE International Conference on Source Code Analysis and Manipulation (SCAM)},
  year      = {2026},
  publisher = {IEEE},
  address   = {Benevento, Italy},
  numpages  = {11},
  url       = {https://arxiv.org/abs/2606.15828}
}

@article{andre_msr,
author = {Robbes, Romain and Matricon, Th{\'e}o and Degueule, Thomas and Hora, Andre and Zacchiroli, Stefano},
title = {Agentic Much? Adoption of Coding Agents on GitHub},
year = {2026},
publisher = {Association for Computing Machinery},
address = {New York, NY, USA},
issn = {1049-331X},
url = {https://doi.org/10.1145/3822180},
doi = {10.1145/3822180},
note = {Just Accepted},
journal = {ACM Trans. Softw. Eng. Methodol.},
month = jun,
numpages = {43}

}

@inproceedings{improta2025qualityinqualityout,
    author = {Improta, Cristina and Tufano, Rosalia and Liguori, Pietro and Cotroneo, Domenico and Bavota, Gabriele},
    title = {Quality In, Quality Out: Investigating Training Data's Role in AI Code Generation},
    year = {2025},
    publisher = {IEEE Press},
    address   = {Piscataway, NJ, USA},
    url = {https://doi.org/10.1109/ICPC66645.2025.00056},
    doi = {10.1109/ICPC66645.2025.00056},
    booktitle = {2025 IEEE/ACM 33rd International Conference on Program Comprehension (ICPC)},
    pages = {454–465},
    numpages = {12}
}

@InProceedings{sun2026evaluatinglanguagemodelscrosslanguage,
  author =	{Sun, Hui and Uch\^{o}a, Anderson and Gheyi, Rohit and Assun\c{c}\~{a}o, Wesley K. G.},
  title =	{{Evaluating Language Models on Cross-Language Code Functional Equivalence}},
  booktitle =	{20th International Symposium on Empirical Software Engineering and Measurement (ESEM 2026)},
  pages =	{41:1--41:21},
  series =	{Leibniz International Proceedings in Informatics (LIPIcs)},
  ISBN =	{978-3-95977-450-5},
  ISSN =	{1868-8969},
  year =	{2026},
  volume =	{394},
  editor =	{Feldt, Robert and Paasivaara, Maria and Mendez, Daniel and Wagner, Stefan and Bar\'{o}n, Marvin Mu\~{n}oz},
  publisher =	{Schloss Dagstuhl -- Leibniz-Zentrum f{\"u}r Informatik},
  address =	{Dagstuhl, Germany},
  URL =		{https://drops.dagstuhl.de/entities/document/10.4230/LIPIcs.ESEM.2026.41},
  URN =		{urn:nbn:de:0030-drops-280091},
  doi =		{10.4230/LIPIcs.ESEM.2026.41}
}

@inproceedings{macedo2024intertransleveragingtransitiveintermediate,
    author = {Macedo, Marcos and Tian, Yuan and Nie, Pengyu and Cogo, Filipe R. and Adams, Bram},
    title = {InterTrans: Leveraging Transitive Intermediate Translations to Enhance LLM-Based Code Translation},
    year = {2025},
    isbn = {9798331505691},
    publisher = {IEEE Press},
    address = {Ottawa, Ontario, Canada},
    url = {https://doi.org/10.1109/ICSE55347.2025.00236},
    doi = {10.1109/ICSE55347.2025.00236},
    booktitle = {Proceedings of the IEEE/ACM 47th International Conference on Software Engineering},
    pages = {1153–1164},
    numpages = {12},
    location = {Ottawa, Ontario, Canada},
    series = {ICSE '25}
}

@inproceedings{HoraFraser2026,
  author    = {Andre Hora and Gordon Fraser},
  title     = {Understanding Bug-Reproducing Tests: A First Empirical Study},
  booktitle = {Proceedings of the 7th ACM/IEEE International Conference on Automation of Software Test (AST)},
  year      = {2026},
  pages     = {93--97},
  publisher = {ACM},
  address   = {New York, NY, USA},
  doi       = {10.1145/3793654.3793752}
}

@article{zhang2025memory,
author = {Zhang, Zeyu and Dai, Quanyu and Bo, Xiaohe and Ma, Chen and Li, Rui and Chen, Xu and Zhu, Jieming and Dong, Zhenhua and Wen, Ji-Rong},
title = {A Survey on the Memory Mechanism of Large Language Model-based Agents},
year = {2025},
issue_date = {November 2025},
publisher = {Association for Computing Machinery},
address = {New York, NY, USA},
volume = {43},
number = {6},
issn = {1046-8188},
url = {https://doi.org/10.1145/3748302},
doi = {10.1145/3748302},
journal = {ACM Trans. Inf. Syst.},
month = sep,
articleno = {155},
numpages = {47}
}

\newpage
\appendix

\renewcommand{\thesection}{Appendix~\Alph{section}}
\renewcommand{\thesubsection}{%
  \Alph{section}.\arabic{subsection}}

\section{LLM Baseline Results}
\label{app:baseline-results}

This appendix complements the baseline comparison presented in
Section~\ref{sec:rq3} with detailed results for the six configurations
evaluated on GWT-Light.
We considered two LLMs, GPT-5-mini and MiniMax-M2.5, each using
zero-shot, one-shot, and Chain-of-Thought (CoT) prompting.
In total, the evaluation comprises 1,026 independent executions
across 171 BDD snapshots.
We report execution outcomes, reproduction recall, and computational
cost for each configuration.

\subsection{Effectiveness Results}
\label{app:baseline-effectiveness}

Table~\ref{tab:app-baseline-effectiveness} presents the execution outcomes
by programming language and overall.
GPT-5-mini achieved its highest success rate with zero-shot, passing
22 out of 171 snapshots (12.9\%), compared with 21 (12.3\%) for
one-shot and 19 (11.1\%) for CoT.
MiniMax-M2.5 passed between 11 (6.4\%) and 19 (11.1\%) snapshots,
with CoT achieving its highest success rate.
Ruby had the lowest success rate across all six configurations,
ranging from 0.0\% to 3.4\%.

Test generation failure was the most frequent outcome in all
six configurations, affecting 71--88 snapshots with GPT-5-mini
across all three prompting strategies and 83--94 with MiniMax-M2.5
under zero-shot and one-shot prompting.

\begin{table*}[!htbp]
  \centering
  \caption{Effectiveness of the baselines by programming language and overall: number of snapshots in each outcome. P: passed; Pt: partial; NP: non-passed; EF: test execution failure; GF: test generation failure (statuses of the heatmap of RQ3). Success is the percentage of passed snapshots overall.}
  \label{tab:app-baseline-effectiveness}
  \setlength{\tabcolsep}{3pt}
  \resizebox{\textwidth}{!}{%
    \begin{tabular}{l|l|rrrrr|rrrrr|rrrrr|rrrrr|r}
      \toprule
                                             &                 & \multicolumn{5}{c}{\textbf{Java} ($n=44$)} & \multicolumn{5}{c}{\textbf{Node.js} ($n=38$)} & \multicolumn{5}{c}{\textbf{Ruby} ($n=89$)} & \multicolumn{5}{c}{\textbf{Overall} ($n=171$)} & \textbf{Success}                                                                                          \\
      \cmidrule(lr){3-7} \cmidrule(lr){8-12} \cmidrule(lr){13-17} \cmidrule(lr){18-22}
      \textbf{Model}                         & \textbf{Prompt} & P                                          & Pt                                            & NP                                         & EF                                             & GF               & P & Pt & NP & EF & GF & P & Pt & NP & EF & GF & P  & Pt & NP & EF & GF & \textbf{(\%)} \\
      \midrule
      \multirow{3}{*}{\textbf{MiniMax-M2.5}} & Zero-shot       & 5                                          & 2                                             & 14                                         & 5                                              & 18               & 6 & 0  & 3  & 15 & 14 & 0 & 2  & 36 & 0  & 51 & 11 & 4  & 53 & 20 & 83 & \textbf{6.4}  \\
                                             & One-shot        & 9                                          & 2                                             & 12                                         & 4                                              & 17               & 6 & 0  & 4  & 17 & 11 & 0 & 1  & 22 & 0  & 66 & 15 & 3  & 38 & 21 & 94 & \textbf{8.8}  \\
                                             & CoT             & 12                                         & 2                                             & 14                                         & 7                                              & 9                & 7 & 0  & 8  & 12 & 11 & 0 & 6  & 38 & 0  & 45 & 19 & 8  & 60 & 19 & 65 & \textbf{11.1} \\
      \midrule
      \multirow{3}{*}{\textbf{GPT-5-mini}}   & Zero-shot       & 10                                         & 4                                             & 10                                         & 2                                              & 18               & 9 & 0  & 0  & 4  & 25 & 3 & 1  & 55 & 0  & 30 & 22 & 5  & 65 & 6  & 73 & \textbf{12.9} \\
                                             & One-shot        & 11                                         & 1                                             & 14                                         & 2                                              & 16               & 8 & 0  & 0  & 5  & 25 & 2 & 0  & 56 & 1  & 30 & 21 & 1  & 70 & 8  & 71 & \textbf{12.3} \\
                                             & CoT             & 8                                          & 2                                             & 12                                         & 4                                              & 18               & 9 & 0  & 0  & 5  & 24 & 2 & 0  & 41 & 0  & 46 & 19 & 2  & 53 & 9  & 88 & \textbf{11.1} \\
      \bottomrule
    \end{tabular}}
\end{table*}

\begin{figure*}[t]
  \centering
  \includegraphics[width=\textwidth]{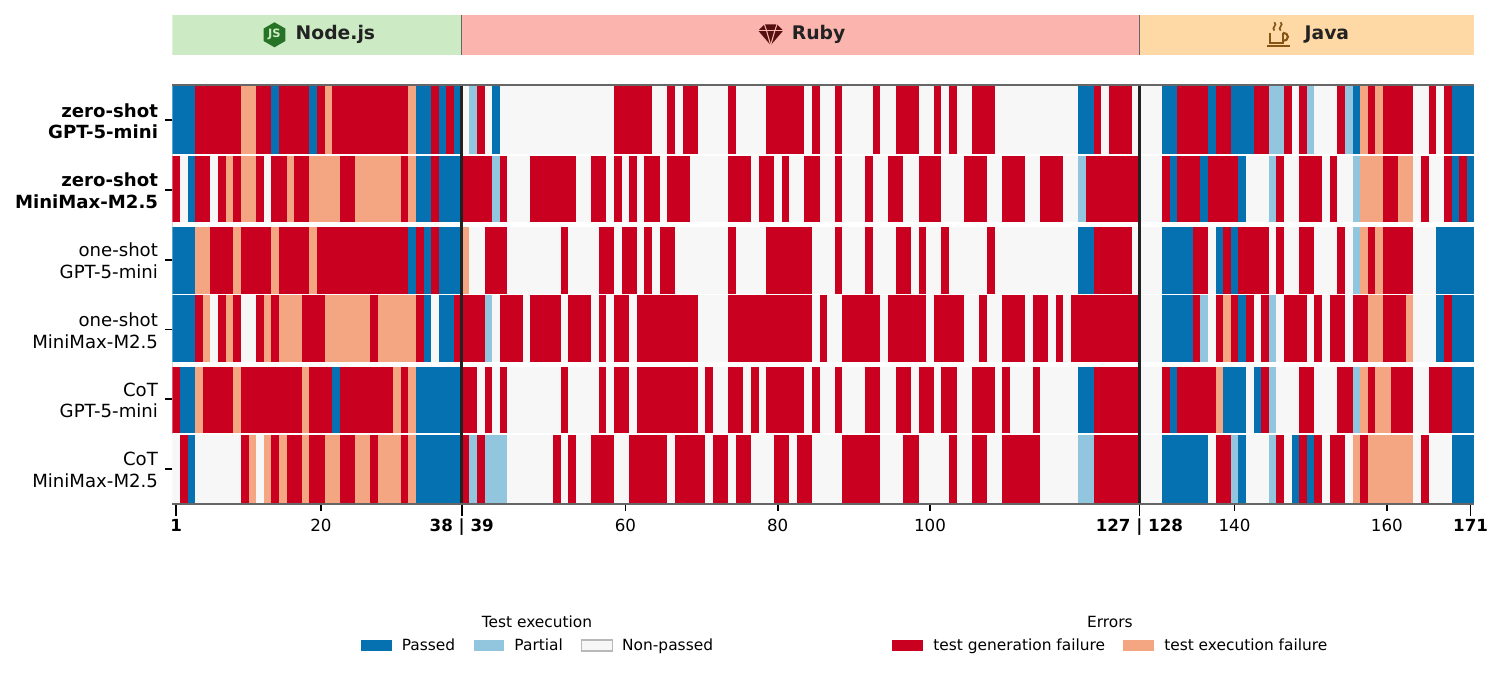}
  \caption{Individual scenario outcomes for the GPT-5-mini and MiniMax-M2.5 baseline models using Zero-shot, one-shot and  Chain-of-Thought prompting.}
  \Description{Heatmap comparing two baseline models across 171 scenarios, grouped by programming language: Node.js, Ruby, and Java. Six rows pair GPT-5-mini and MiniMax-M2.5 for each prompting strategy. Zero-shot appears first and is emphasized as the primary comparison, followed by one-shot and Chain-of-Thought.}
  \label{fig:baseline-scenarios}
\end{figure*}

\subsection{Similarity Results}
\label{app:baseline-similarity}

Table~\ref{tab:app-baseline-similarity} summarizes reproduction recall
for executions with a recorded value, ranging from 56 to 92 per
configuration.
MiniMax-M2.5 obtained average recall values between 0.29 and 0.32,
compared with 0.25 to 0.31 for GPT-5-mini.
Similarly, the median ranged from 0.28 to 0.33 for MiniMax-M2.5 and
from 0.18 to 0.25 for GPT-5-mini; the first quartile was zero for
GPT-5-mini under all three prompting strategies.
One-shot yielded the highest average recall for MiniMax-M2.5 and
CoT for GPT-5-mini, even though zero-shot achieved the highest
functional success rate for GPT-5-mini.
These results reinforce that reproduction recall measures agreement
with reference method calls rather than functional correctness.

\begin{table}[!htbp]
  \centering
  \caption{Reproduction recall of the baselines. N is the number of executions with a recorded recall value, excluding LLM and infrastructure failures.}
  \label{tab:app-baseline-similarity}
  \begin{tabular}{llr|r|rrrrr}
    \toprule
    \textbf{Model}                         & \textbf{Prompt} & \textbf{N} & \textbf{Average} & \textbf{Min} & \textbf{Q1} & \textbf{Median} & \textbf{Q3} & \textbf{Max} \\
    \midrule
    \multirow{3}{*}{\textbf{MiniMax-M2.5}} & Zero-shot       & 68         & 0.32             & 0.00         & 0.09        & 0.28            & 0.46        & 1.00         \\
                                           & One-shot        & 56         & 0.32             & 0.00         & 0.10        & 0.33            & 0.50        & 1.00         \\
                                           & CoT             & 87         & 0.29             & 0.00         & 0.06        & 0.29            & 0.50        & 0.90         \\
    \midrule
    \multirow{3}{*}{\textbf{GPT-5-mini}}   & Zero-shot       & 92         & 0.28             & 0.00         & 0.00        & 0.25            & 0.39        & 1.00         \\
                                           & One-shot        & 92         & 0.25             & 0.00         & 0.00        & 0.18            & 0.39        & 1.00         \\
                                           & CoT             & 74         & 0.31             & 0.00         & 0.00        & 0.25            & 0.50        & 1.00         \\
    \bottomrule
  \end{tabular}
\end{table}

\subsection{Computational Cost}
\label{app:baseline-cost}

Table~\ref{tab:app-baseline-full} presents the distribution of financial
cost, input tokens, output tokens, and reproduction recall for each baseline
configuration, including the minimum, quartiles, maximum, and mean.
Across the 171 snapshots, total recorded costs ranged from US\$0.89 to US\$0.96
for GPT-5-mini and from US\$0.41 to US\$0.50 for MiniMax-M2.5.
Zero-shot achieved the lowest cost per successful snapshot for GPT-5-mini
(US\$0.0408), while CoT achieved the lowest for MiniMax-M2.5 (US\$0.0214).
The highest cost per success was observed for GPT-5-mini with
CoT prompting (US\$0.0505).

Median input consumption ranged from 1,868 to 2,550 tokens across
configurations, while median output consumption was higher for
GPT-5-mini (1,882--2,269 tokens) than for MiniMax-M2.5 (0--769 tokens).
Truncated executions represented 17.0--26.9\% of GPT-5-mini runs and
28.7--49.7\% of MiniMax-M2.5 runs.
These executions recorded zero output tokens despite having a recorded
financial cost in most cases; therefore, output-token distributions
include executions with recorded zero output.

\begin{table*}[!htbp]
  \centering
  \caption{Distribution of cost, tokens, and reproduction recall per execution for each baseline configuration. Output tokens consider only executions with recorded output; recall considers only executions with a recorded value and no LLM or infrastructure failure.}
  \label{tab:app-baseline-full}
  \resizebox{\textwidth}{!}{%
    \begin{tabular}{lll|r|rrrrr}
      \toprule
      \textbf{Model}                          & \textbf{Prompt}            &                                & \textbf{Average} & \textbf{Min} & \textbf{Q1} & \textbf{Median} & \textbf{Q3} & \textbf{Max} \\
      \midrule
      \multirow{12}{*}{\textbf{MiniMax-M2.5}} & \multirow{4}{*}{Zero-shot} & \textbf{Financial Cost (US\$)} & 0.002639         & 0.000819     & 0.001517    & 0.002842        & 0.003678    & 0.004590     \\
                                              &                            & \textbf{Input Tokens}          & 2,136.11         & 842          & 1,539.5     & 1,868           & 2,747       & 5,643        \\
                                              &                            & \textbf{Output Tokens}         & 738.91           & 0            & 0           & 733             & 1,041       & 3,002        \\
                                              &                            & \textbf{Recall}                & 0.32             & 0.00         & 0.09        & 0.28            & 0.46        & 1.00         \\
      \cmidrule(l){2-9}
                                              & \multirow{4}{*}{One-shot}  & \textbf{Financial Cost (US\$)} & 0.002963         & 0.000888     & 0.001884    & 0.003453        & 0.003899    & 0.004604     \\
                                              &                            & \textbf{Input Tokens}          & 2,184.32         & 1,015        & 1,768.5     & 2,047           & 2,601.5     & 4,092        \\
                                              &                            & \textbf{Output Tokens}         & 604.11           & 0            & 0           & 0               & 1,015       & 3,000        \\
                                              &                            & \textbf{Recall}                & 0.32             & 0.00         & 0.10        & 0.33            & 0.50        & 1.00         \\
      \cmidrule(l){2-9}
                                              & \multirow{4}{*}{CoT}       & \textbf{Financial Cost (US\$)} & 0.002374         & 0.000694     & 0.001399    & 0.001920        & 0.003694    & 0.004671     \\
                                              &                            & \textbf{Input Tokens}          & 2,600.79         & 1,417        & 2,134.5     & 2,503           & 2,969.5     & 4,329        \\
                                              &                            & \textbf{Output Tokens}         & 744.00           & 0            & 0           & 769             & 1,036       & 3,002        \\
                                              &                            & \textbf{Recall}                & 0.29             & 0.00         & 0.06        & 0.29            & 0.50        & 0.90         \\
      \midrule
      \multirow{12}{*}{\textbf{GPT-5-mini}}   & \multirow{4}{*}{Zero-shot} & \textbf{Financial Cost (US\$)} & 0.005256         & 0.001485     & 0.004287    & 0.006002        & 0.006223    & 0.007108     \\
                                              &                            & \textbf{Input Tokens}          & 2,336.94         & 819          & 1,755.5     & 2,171           & 2,712.5     & 6,143        \\
                                              &                            & \textbf{Output Tokens}         & 1,916.39         & 0            & 1,192.5     & 2,269           & 2,965.5     & 3,000        \\
                                              &                            & \textbf{Recall}                & 0.28             & 0.00         & 0.00        & 0.25            & 0.39        & 1.00         \\
      \cmidrule(l){2-9}
                                              & \multirow{4}{*}{One-shot}  & \textbf{Financial Cost (US\$)} & 0.005367         & 0.001363     & 0.004243    & 0.006034        & 0.006646    & 0.007158     \\
                                              &                            & \textbf{Input Tokens}          & 2,459.76         & 1,154        & 2,015       & 2,469           & 2,753       & 4,633        \\
                                              &                            & \textbf{Output Tokens}         & 1,861.41         & 0            & 1,055       & 2,106           & 3,000       & 3,000        \\
                                              &                            & \textbf{Recall}                & 0.25             & 0.00         & 0.00        & 0.18            & 0.39        & 1.00         \\
      \cmidrule(l){2-9}
                                              & \multirow{4}{*}{CoT}       & \textbf{Financial Cost (US\$)} & 0.005616         & 0.001545     & 0.004574    & 0.006586        & 0.006682    & 0.007215     \\
                                              &                            & \textbf{Input Tokens}          & 2,602.84         & 1,381        & 2,196       & 2,550           & 2,905       & 4,861        \\
                                              &                            & \textbf{Output Tokens}         & 1,661.75         & 0            & 0           & 1,882           & 3,000       & 3,000        \\
                                              &                            & \textbf{Recall}                & 0.31             & 0.00         & 0.00        & 0.25            & 0.50        & 1.00         \\
      \bottomrule
    \end{tabular}}
\end{table*}

\end{document}